\documentclass{aa}  

\usepackage{graphicx}
\usepackage{txfonts}
\newcommand{\ho}{{\ensuremath{\mathrm{H}_0}}}
\newcommand{\teff}{\ensuremath{\mathrm{T}_\mathrm{eff}}}
\newcommand{\vturb}{\ensuremath{\mathrm{v}_\mathrm{turb}}}
\newcommand{\logg}{\ensuremath{\mathrm{\log(g)}}}
\newcommand{\fe}{\ensuremath{\mathrm{Fe}}}
\newcommand{\ox}{\ensuremath{\mathrm{O}}}
\newcommand{\feh}{\ensuremath{\mathrm{[Fe/H]}}}
\newcommand{\oxh}{\ensuremath{\mathrm{[O/H]}}}

\newcommand{\mi}[2]{\ensuremath{\mathrm{m}_{\mathrm{#1}}^{\mathrm{#2}}}}

\newcommand{\lcdm}{\ensuremath{\Lambda\mathrm{CDM}}}
\newcommand{\sne}{supernov\ae}
\newcommand{\kmsmpc}{\ensuremath{\mathrm{km/s/Mpc}}}

\begin{document} 

   \title{The iron and oxygen content of LMC Classical Cepheids and its implications for the extragalactic distance scale and Hubble constant\thanks{Based on observations collected at the European Southern Observatory under ESO programmes 66.D-0571 and 106.21ML.003.\\
   Tables 1 to 8 and the content of Appendix B are only available in electronic form at the CDS via anonymous ftp to cdsarc.u-strasbg.fr (130.79.128.5) or via http://cdsweb.u-strasbg.fr/cgi-bin/qcat?J/A+A/
   }}
  \titlerunning{The iron and oxygen content of LMC Classical Cepheids}

   \subtitle{Equivalent width analysis with Kurucz stellar atmosphere models}

   \author{Martino Romaniello \inst{1}
            \and
           Adam Riess \inst{2,3}
           \and
           Sara Mancino \inst{1}
           \and
           Richard I. Anderson \inst{4}
           \and
           Wolfram Freudling \inst{1}
           \and
           Rolf-Peter Kudritzki\inst{5,6}
           \and
           Lucas Macr\`i \inst{7}
           \and
           Alessio Mucciarelli \inst{8,9}
           \and
           Wenlong Yuan \inst{3}
          }

   \institute{European Southern Observatory, Karl-Schwarzschild-Strasse 2, D-85478 Garching bei M\"unchen\\
              \email{martino.romaniello@eso.org}
              \and
    Space Telescope Science Institute, 3700 San Martin Drive, Baltimore, MD 21218, USA
        \and
    Department of Physics and Astronomy, Johns Hopkins University, Baltimore, MD 21218, USA
        \and
    Institute of Physics, Laboratory of Astrophysics, \'Ecole Polytechnique F\'ederale del Lausanne (EPFL), Observatoire de Suverny, 1290 Versoix, Switzerland
        \and
        LMU M\"unchen, Universit\"atssternwarte, Scheinerstrasse 1, D-81679, Germany
        \and
        Institute for Astronomy, University of Hawaii at Manoa, 2680 Woodlawn Drive, Honolulu, HI 96822, USA
        \and
    Mitchell Institute for Fundamental Physics \& Astronomy, Department of Physics \& Astronomy, Texas A\&M University, College Station, TX 77843, USA
        \and
       Dipartimento di Fisica e Astronomia Augusto Righi, Università degli Studi di Bologna, Via Gobetti 93/2, 40129 Bologna, Italy
       \and
       INAF – Osservatorio di Astrofisica e Scienza dello Spazio di Bologna, Via Gobetti 93/3, 40129 Bologna, Italy
    }

   \date{Received September 15, 1996; accepted March 16, 1997}

 
  \abstract
   {Context. Classical Cepheids are primary distance indicators and a crucial stepping stone in determining the present-day value of the Hubble constant \ho\ to the precision and accuracy required to constrain apparent deviations from the \lcdm\ Concordance Cosmological Model.}
   {Aims. We measured the iron and oxygen abundances of a statistically significant sample of 89 Cepheids in the Large Magellanic Cloud (LMC), one of the anchors of the local distance scale, quadrupling the prior sample and including 68 of the 70 Cepheids used to constrain \ho\ by the SH0ES program. The goal is to constrain the extent to which the luminosity of Cepheids is influenced by their chemical composition, which is an important contributor to the uncertainty on the determination of the Hubble constant itself and a critical factor in the internal consistency of the distance ladder.}
   {We derived stellar parameters and chemical abundances from a self-consistent spectroscopic analysis based on equivalent width of absorption lines.}
   {The iron distribution of Cepheids in the LMC can be very accurately described by a single Gaussian with a mean $\overline{\feh}=-0.409\pm0.003$~dex and $\sigma=0.076\pm0.003$~dex. We estimate a systematic uncertainty on the absolute mean values of 0.1~dex. The width of the distribution is fully compatible with the measurement error and supports the low dispersion of 0.069~mag seen in the near-infrared Hubble Space Telescope LMC period--luminosity relation. The uniformity of the abundance has the important consequence that the LMC Cepheids alone cannot provide any meaningful constraint on the dependence of the Cepheid period--luminosity relation on chemical composition at any wavelength. This revises a prior claim based on a small sample of 22 LMC Cepheids that there was little dependence (or uncertainty) between composition and near-infrared luminosity, a conclusion which would produce an apparent conflict between anchors of the distance ladder with different mean abundance.
   The chemical homogeneity of the LMC Cepheid population makes it an ideal environment in which to calibrate the metallicity dependence between the more metal-poor Small Magellanic Cloud and metal-rich Milky Way and NGC~4258.}
   {}


   \maketitle
%

\section{Introduction} \label{Sec:intro}
    The current discrepancy regarding the present-day value of the Hubble constant \ho\ is arguably one of the most far-reaching open problems in astrophysics today. The extrapolation of early-Universe cosmic microwave background measurements assuming the \lcdm\ Cosmological Concordance Model predicts a value of $\ho=67.4\pm0.5\ \kmsmpc$ \citep{planck2020}. Comparison to the best late-Universe local measurement of $\ho=73.2\pm1.3\ \kmsmpc$ \citep[][SH0ES project]{riess2021}, which is independent of any such assumptions, reveals a significant $4.2\sigma$ tension. Moreover, {all} recent local determinations, obtained by different groups with different methods, exceed the early-Universe prediction \citep[e.g.][and references therein]{verde2019, riess2020}. The conclusion that there is indeed a meaningful difference between early- and local-Universe values of \ho\ is therefore by now almost inescapable. This is hugely consequential: in spite of its many phenomenological successes, the \lcdm\ model does not provide a complete description of the Universe

    Classical Cepheids play a pivotal role in this context, because they are primary distance indicators and a crucial stepping stone in determining the present-day value of the Hubble constant to the precision and accuracy required to constrain possible deviations from \lcdm. In particular, the SH0ES project has built a clean, three-rung distance ladder based on them and on type-Ia \sne. Their goal is to measure the local value of the Hubble constant \ho\ to an accuracy of better than 1\% so as to provide a meaningful match to the early-Universe value derived from the cosmic microwave background.
    
     Cepheids can be used to measure distances because their brightness changes periodically with time according to the Leavitt Law \citep{leavitt1912}, which links the period of such pulsations, which is independent of distance, to their mean magnitude, which scales as the inverse of the distance squared (period--luminosity (PL) relation). The uncertainty on the behaviour of the PL relation with chemical composition is a major contributor to the overall SH0ES \ho\ error budget \citep[0.9\% out of 1.8\%,][]{riess2021}. Indeed, recent results using Galactic and Magellanic Cloud Cepheids confirm that, while progress is being made in terms of sample sizes, as crucially enabled by GAIA parallaxes, the slope of a potential dependence of the Cepheid luminosity on metal content remains too loosely constrained, with significant differences between different studies; for example $-0.221\pm0.051$~mag/dex \cite[][see also \citealt{gieren2018}]{breuval2021} versus $-0.456\pm0.099$~mag/dex \cite{ripepi2021} for the slope in the $\mathrm{K}_\mathrm{s}$ band.

    Here, we provide accurate direct abundance measurements from high-resolution spectroscopy of Cepheids stars in the Large Magellanic Cloud (LMC), one of the three anchor galaxies in the SH0ES distance ladder \citep{riess2019}. The paper is organised as follows. In section~\ref{Sec:obs_reduction} we present the data and their reduction to calibrated one-dimensional spectra. Section~\ref{Sec:params} describes the derivation of the stellar parameters that are needed in order to measure chemical abundances. These are discussed in Sections~\ref{Sec:Fe_abus} and \ref{Sec:O_abus} for iron and oxygen, respectively. Finally, conclusions are drawn in Section~\ref{Sec:conclusions}.

\section{Observations and data reduction} \label{Sec:obs_reduction}
    We assembled our sample of {89} Classical Cepheids in the LMC by combining proprietary data for {67} out of the 70 stars that define the SH0ES fiducial PL relation \citep{riess2019} with 22 archival spectra observed in the same instrumental setup. For the mentioned sample of 67 stars, this is the first measurement of their chemical composition. Iron abundances for the other ones were published by \citet[][OGLE510/HV879 is in common between the two samples and is considered here only once; see Table~\ref{Tab:obslog}]{romaniello2008}. Here, we revise those measurements, which we find were affected by undetected systematic errors.
    
\begin{figure*}
   \centering
   \includegraphics[scale=0.5]{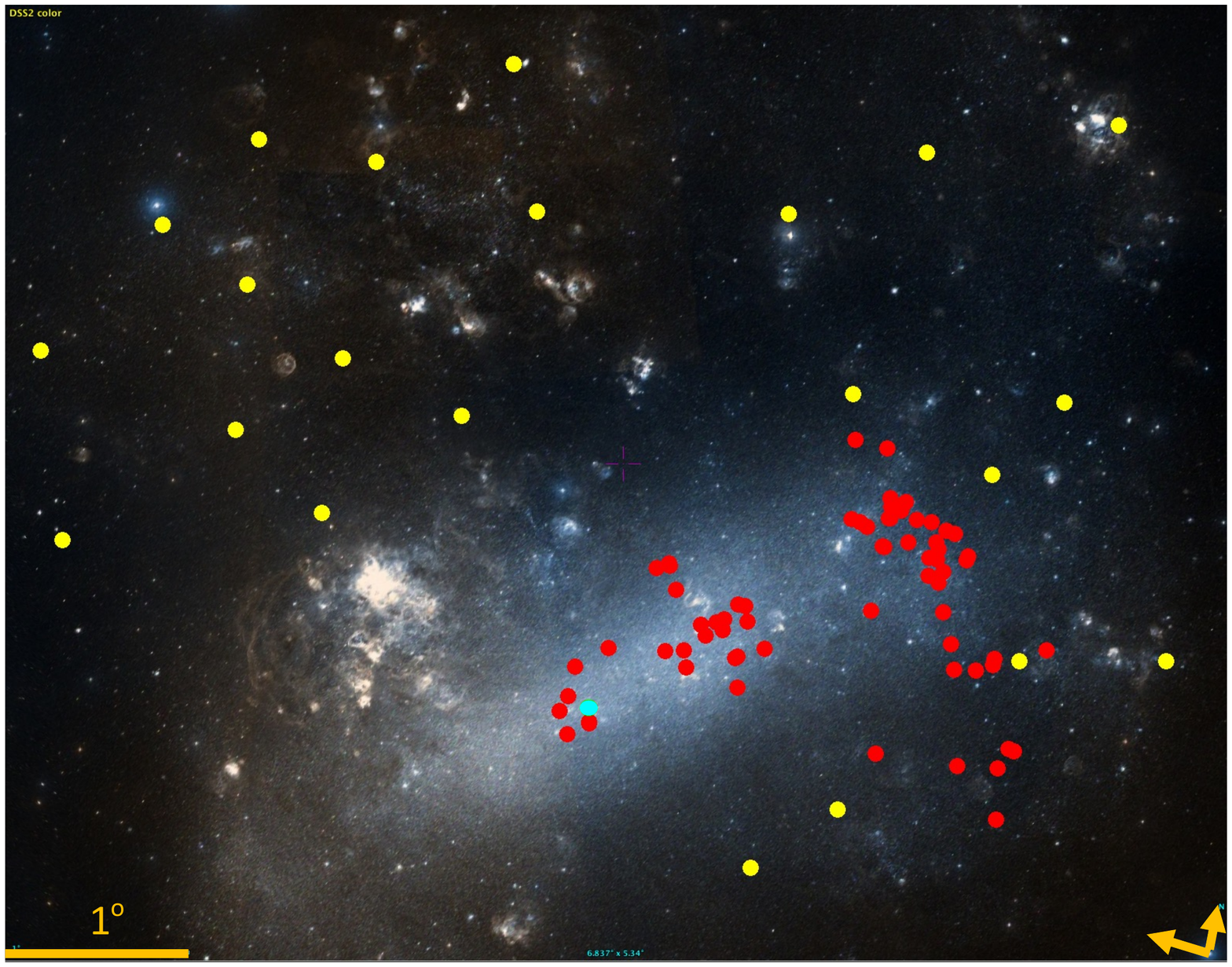}
   \caption{Distribution on the plane of the sky of the 68 SH0ES Classical Cepheids targeted in this paper (red dots). The two stars for which we do not have spectroscopic data are marked in cyan. The stars from the \cite{romaniello2008} sample that we re-analyse here are shown as yellow dots. The figure was generated with the `Aladin sky atlas' software \citep{aladin} using a DSS2 colour image of the LMC as background.}
              \label{Fig:onsky}
    \end{figure*}

\begin{figure}
   \centering
   \includegraphics[scale=0.3]{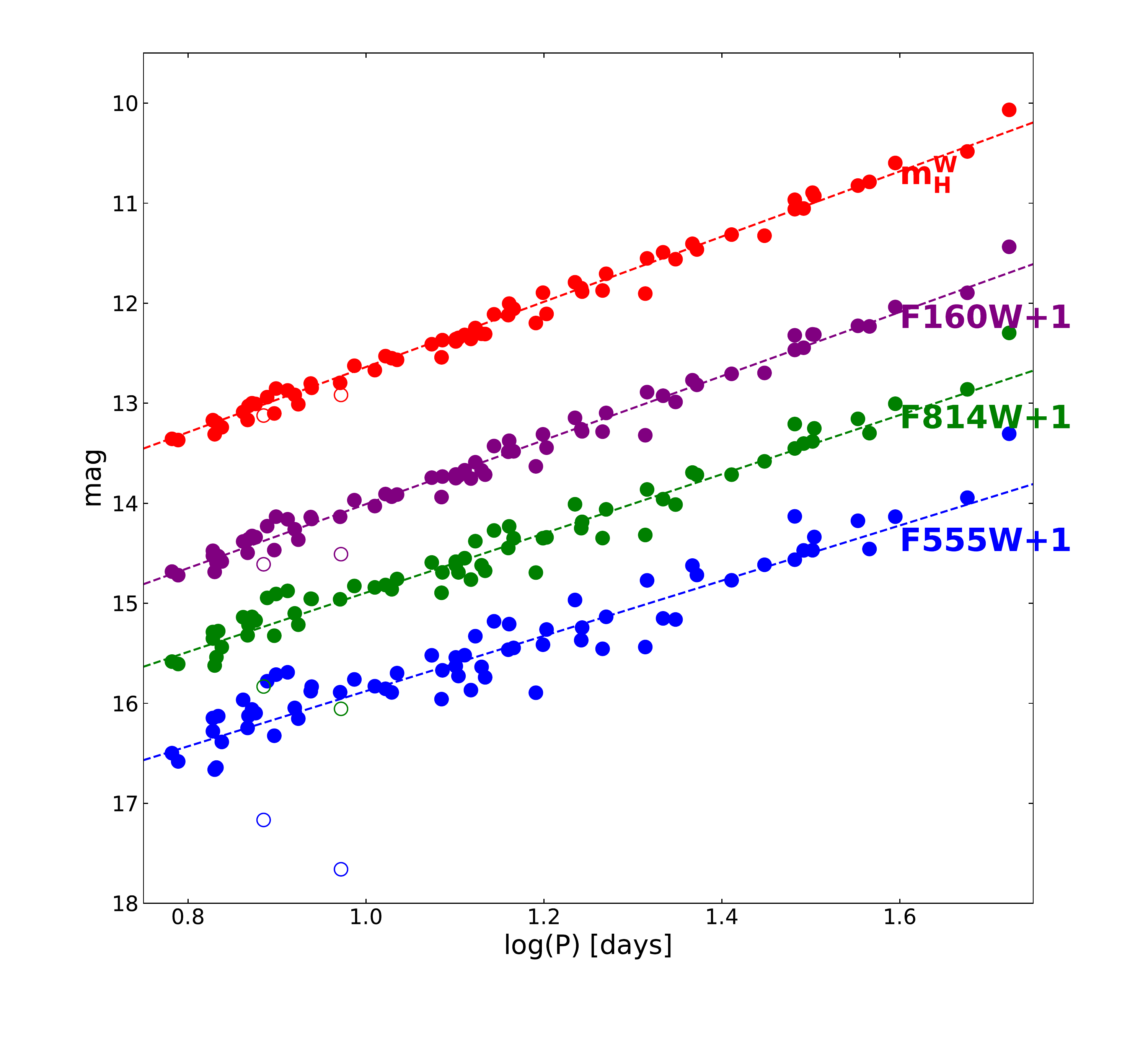}
   \caption{PL relations in the various bands of the sky of the SH0ES Classical Cepheids \citep[see Table~2 and Figure~3]{riess2019}. The two for which we do not have spectroscopic data are marked as open circles. The scatter in the \mi{H}{W} relation is a mere 0.075~mag, making it the tightest to date.}
    \label{Fig:PL}
    \end{figure}

    All of the stars are listed in the OGLE IV catalogue \citep{udalski2015, soszynski2015}. In the following, we use their OGLE identifier, which we shortened for convenience (e.g. OGLE-LMC-CEP-0966 to OGLE0966). Their on-sky distribution and the PL relations in the various bands of the SH0ES sample are presented in Figures~\ref{Fig:onsky} and \ref{Fig:PL}, respectively.
    
    We have excluded two of the SH0ES stars from our spectroscopic sample (OGLE1940, OGLE1945; open circles in Figure~\ref{Fig:PL}) because they are too faint in the optical to obtain spectra in a reasonable amount of time. They are also significant outliers from the PL relation in all magnitudes except \mi{H}{W} and excluding them from the analysis has no impact on our conclusions.    

    All {89} stars were observed with the UVES spectrograph at the ESO VLT telescope in the UVES~580 setting. This spectrograph covers the wavelength range between 4780 and 6800~\AA\ with a gap between 5760 and 5830~\AA. The instrumental resolving power is $R\sim50,000$, but Cepheids have intrinsically broad spectral lines, yielding an effective value of $R\sim20,000$. The proprietary observations of the SH0ES sample were carried out in Service Mode. Observing constraints and exposure times were adapted to each star in order to deliver data of uniform quality (signal-to-noise ratios higher than 40). No constraints were set on the pulsational phase at the time of observation, which is therefore random. In two cases (OGLE0545, OGLE1647), observations were at first executed outside of the constraints we had specified. They were subsequently successfully repeated and we did not use the first set of data, because they did not add significantly to the final quality. The log of the observations we did use in the analysis is reported in Table~\ref{Tab:obslog}. We refer the reader to \cite{romaniello2008} for the details of the observations of the archival Cepheids.

    \begin{table}
        \caption[]{Log of the spectroscopic observations of the 68 SH0ES Cepheids.}
        \begin{center}
        \begin{tabular}{ccc}
            \hline
            Cepheid &  Filename timestamp\tablefootmark{a} &  MJD$\tablefootmark{b}$ \\
            \hline
            OGLE0434 & 2020-11-07T02:47:44.709 & 59160.11649 \\ 
OGLE0434 & 2020-11-07T02:47:52.206 & 59160.11658 \\ 
OGLE0501 & 2020-10-24T04:50:05.197 & 59146.20145 \\ 
OGLE0501 & 2020-10-24T04:50:13.252 & 59146.20154 \\ 
OGLE0512 & 2020-10-24T05:25:26.486 & 59146.22600 \\ 
OGLE0512 & 2020-10-24T05:25:36.488 & 59146.22612 \\ 
OGLE0528 & 2020-10-09T07:28:41.908 & 59131.31160 \\ 
OGLE0528 & 2020-10-09T07:28:48.852 & 59131.31168 \\ 
OGLE0545 & 2020-11-07T03:00:20.624 & 59160.12524 \\ 
OGLE0545 & 2020-11-07T03:00:29.841 & 59160.12535 \\ 
\ldots & \ldots & \ldots \\ 
\hline
OGLE0510 & 2000-10-11T04:34:30.713 & 51828.19063 \\ 
OGLE0510 & 2000-10-11T04:44:53.443 & 51828.19784 \\ 

            \hline
        \end{tabular}
        \end{center}
        \tablefoot{\\This table is available in its entirety in machine-readable form.\\
        All stars were observed in the same instrumental setting: UVES 580. Data for OGLE0510 was already available in the ESO Science Archive, so we retrieved it from there and analysed it in the same way as the others. Observations longer than 45 minutes were split in two back-to-back exposures.\\
        \tablefoottext{a}{The name of the raw science file in the ESO Science Archive is UVES.timestamp.}\\
        \tablefoottext{b}{$\mathrm{MJD} = \mathrm{JD} - 2,400,000.5.$}}
        \label{Tab:obslog}
    \end{table}

    We downloaded the raw science files from the ESO Science Archive\footnote{\url{http://archive.eso.org}} both for our proprietary data and the archival ones, together with the calibrations provided by the system according to the instrument Calibration Plan \citep{uves_calplan}. We reduced the data with the instrument pipeline version 6.1.3 \citep{uves_pipeman}, executed within the ESOReflex environment \citep{reflex}, and added a custom step to the default processing cascade to combine repeated observations of the same targets. In all cases, these are taken in sequence (see Table~\ref{Tab:obslog}), so a simple co-addition of the raw 2D frames is sufficient.
    
    In the first pass, all data were processed with the default pipeline recipe parameters, which were then optimised by individually inspecting the results. In all cases, we reduced the rejection threshold during spectral extraction from 10 to $5\sigma$ (parameter reduce.extract.kappa in the uves\_obs\_scired recipe) to limit the impact of cosmic ray hits and detector defects. Visual inspection of the results confirmed that no significant residuals are present. In the case of star OGLE0936, the trace of  a bright neighbouring star is clearly visible in the 2D spectrum and the default extraction window includes them both. We therefore tailored the window to only include the star of interest (parameter reduce.extract.kappa=26 in the uves\_obs\_scired recipe). In all other cases, the default recipe parameters were confirmed to be adequate and were left unchanged.
    
\section{Stellar parameters} \label{Sec:params}
    \subsection{The equivalent width method} \label{Sec:params_derivation}
    In order to self-consistently determine the stellar parameters (effective temperature \teff, surface gravity in logarithmic units \logg\ and microturbulent velocity \vturb) and chemical abundances, we use the equivalent width (EW) method. Very briefly, for each star, the EWs of iron lines as measured in the observed spectrum are compared to the ones from a stellar atmosphere model for a given set of stellar parameters to derive the abundances from each individual line. These are then iterated upon until convergence is reached when the following conditions are met:
    
    \begin{itemize}
        \item {\it Effective temperature} is derived by imposing excitation equilibrium, that is, by imposing that there be no residual correlation between the iron abundance and the excitation potential $\chi$ of the neutral iron lines. As demonstrated by \cite{mucciarelli2020} for example, above a metallicity of $\sim-1.5$, spectroscopic temperatures provide an unbiased estimate.
        \item The best value of {\it surface gravity} comes from imposing ionisation equilibrium, thus requiring that, for a given species, the abundance is the same within the uncertainties from lines of two different ionisation states (in our case, neutral and singly ionised iron lines).
        \item {\it Microturbulent velocity} is set by requiring that there be no residual correlation between the iron abundance and the line EW as a measure of the line strength.
        \item The final {\it iron abundance} is the mean of the iron abundances from the individual lines in the convergence iteration.
        \item The initial values of the parameters were set for all stars as follows: $\teff=5500$~K, $\logg=1~\mathrm{cm}/\mathrm{s}^2$, $\feh=-0.33$~dex and $\vturb=2.50$~km/s. Changing each of them by up to a factor of two influences the resulting abundances randomly at the level of a few hundredths of a dex. 
        \item The {\it abundance of oxygen} is then measured with the stellar parameters determined as described above. 
    \end{itemize}
    
    We used the list provided by \cite{genovali2013}  as input for the spectral location of unblended FeI and FeII lines (Table~\ref{Tab:linelist}); it was compiled specifically for Cepheid stars, thus providing a clean set of lines to avoid line crowding  as much as possible for these types of stars, which have intrinsically broad features (FWHM$\sim$0.2-0.3~\AA). We measured the EWs with the DAOSPEC code \citep{daospec}, which was executed on our entire sample of stars using the 4DAO software \citep{4dao}.
    
    \begin{table}
        \caption[]{List of the 189 FeI, 28 FeII, and 2 OI lines used as input in the analysis.}
        \begin{center}
        \begin{tabular}{cccc}
            \hline
            $\lambda$ [\AA] &  Ion &  log(gf)\tablefootmark{a} & EP [eV]\tablefootmark{a}\\
            \hline
            4892.86 & FeI & -0.876 & 4.22 \\
4893.81 & FeII & -4.157 & 2.83 \\
4917.23 & FeI & -1.160 & 4.19 \\
4923.92 & FeII & -1.559 & 2.89 \\
4924.77 & FeI & -2.114 & 2.28 \\
4932.08 & FeI & -1.483 & 4.65 \\
4950.10 & FeI & -1.500 & 3.42 \\
4973.10 & FeI & -0.690 & 3.96 \\
4993.35 & FeII & -3.684 & 2.81 \\
4994.13 & FeI & -3.080 & 0.92 \\
\ldots & \ldots & \ldots & \ldots \\ 

            \hline
        \end{tabular}
        \end{center}
        \tablefoot{\\This table is available in its entirety in machine-readable form.\\
        From left to right the columns display wavelength, ion identification, oscillator strength log(gf) and excitation potential EP values \citep[adapted from][with updated values from the Kurucz/Castelli database \tablefootmark{a}]{genovali2013}. As part of the abundance analysis, individual lines may be rejected based on the criteria detailed in Section~\ref{Sec:params_derivation}. The distributions of the lines retained in the analysis are shown as histograms in Figure~\ref{Fig:nlines}.}
        \tablefoottext{a}{\url{https://wwwuser.oats.inaf.it/castelli/linelists.html}}
        \label{Tab:linelist}
    \end{table}

\begin{figure}
    \centering
    \includegraphics[scale=0.4]{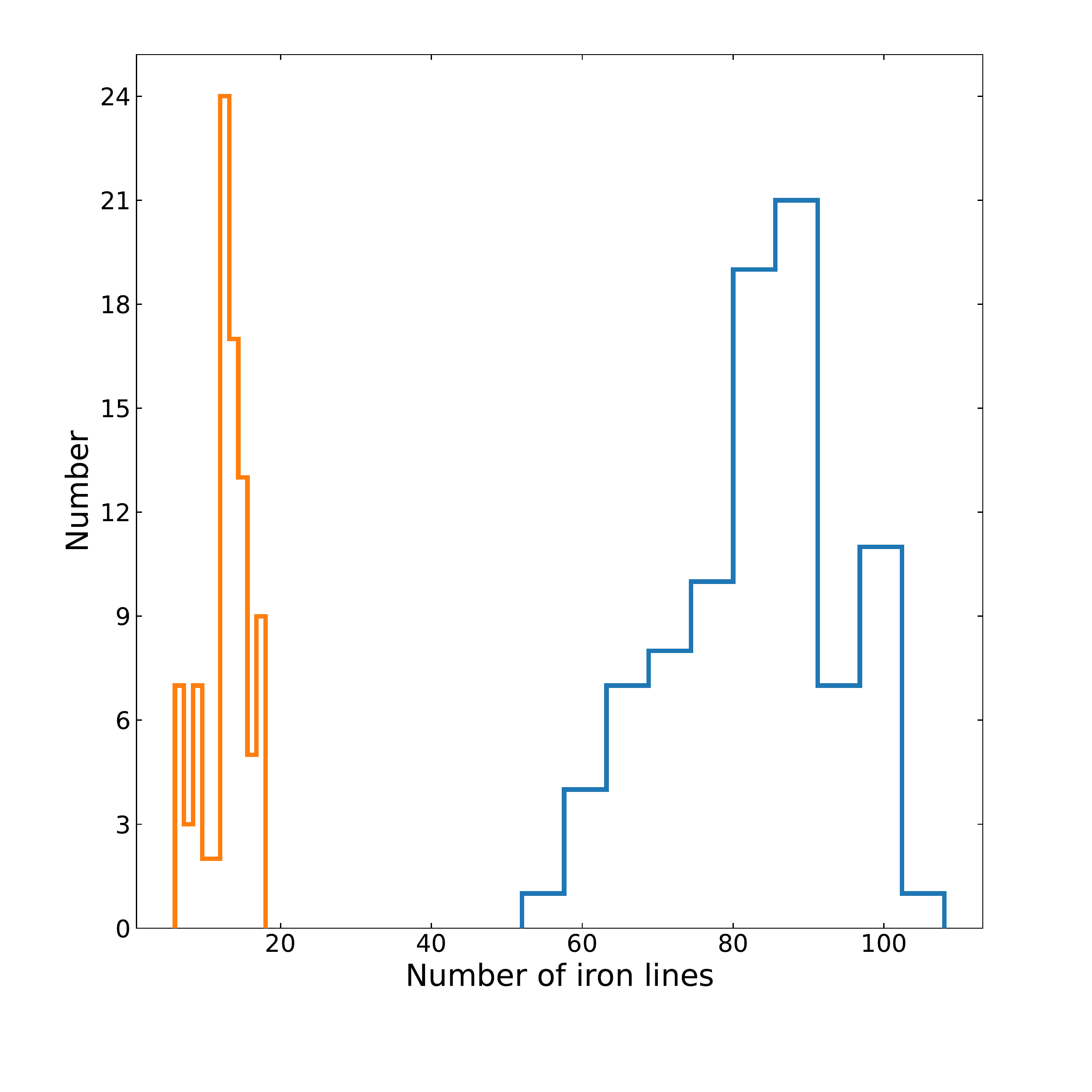}
    \caption{Number of FeI (blue histogram) and FeII (orange histogram) spectral lines retained in the abundance analysis of each of the { 89} programme stars after outlier rejection as part of the fitting procedure with the GALA code (see Section~\ref{Sec:params_derivation}). The full input line list is reported in Table~\ref{Tab:linelist}.}
    \label{Fig:nlines}
    \end{figure}

    In our analysis, we adopted the GALA implementation of the EW method, which was extensively and specifically tested on UVES data \citep{gala}. As input physics, we used  the classical grid of ATLAS9 Local Thermodynamical Equilibrium (LTE) stellar atmosphere models \citep{atlas9}. During the analysis, for each spectrum, lines are rejected from the fit based on a number of quality criteria, so that the final set of lines used in the analysis varies from one spectrum to another, depending on circumstances, such as signal-to-noise ratio and line depth. However, in all cases, a sufficient number of lines are retained such that the resulting solution is meaningful in terms of stellar parameters and elemental abundances. This is shown in Figure~\ref{Fig:nlines}, where the number of iron lines used in the analysis for each spectrum is given.

    Once convergence is reached, the uncertainty on the elemental abundance is computed, taking into account the covariance among the stellar parameters according to the prescription of \cite{cayrel2004}.
    
\begin{figure*}
    \resizebox{\hsize}{!}
     {\includegraphics[]{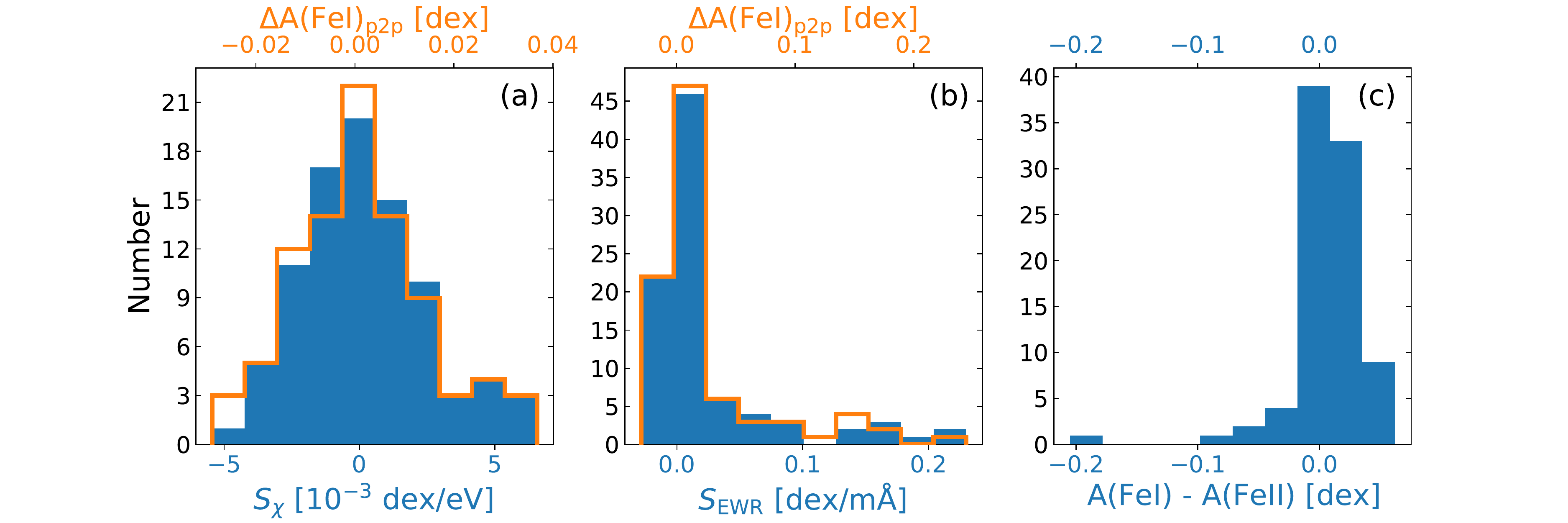}}
      \caption{Histograms of the output values of the diagnostics used to derive the stellar parameters for the { 89} stars in our combined sample.
      {\it Panel~(a):} Slope $S_\chi$ of FeI abundance vs. excitation potential $\chi$ used to derive the effective temperature \teff (solid blue curve, bottom x scale). The corresponding peak-to-peak values of the spread in iron abundance ($\Delta\mathrm{A(FeI)}_\mathrm{p2p}$, see Equation~(\ref{Eq:p2p})) are plotted as an open orange curve (top x scale).
      {\it Panel~(b):} Slope $S_\mathrm{EWR}$ of the relation between FeI abundance and reduced EW $\mathrm{EWR}=\log(\mathrm{EW}/\lambda)$, which ranges between $-5.5$ and $-4.5$, and  is used to constrain the microturbulent velocity \vturb (solid blue curve, bottom x scale). Also here, the corresponding peak-to-peak values of the spread in iron abundance ($\Delta\mathrm{A(FeI)}_\mathrm{p2p}$, see Equation~(\ref{Eq:p2p})) are plotted as an open orange curve (top x scale).
      {\it Panel~(c):} Difference between FeI and FeII, the minimisation of which is used to determine the surface gravity $\logg$. The mean and standard deviation of the distribution are 0.007 and 0.033~dex, respectively, indicating excellent agreement between the derived FeI and FeII abundances, and hence gravity, with no appreciable systematic errors.}
         \label{Fig:gala_diagnostics}
    \end{figure*}

\begin{figure*}
        \centering
        \includegraphics[scale=0.4]{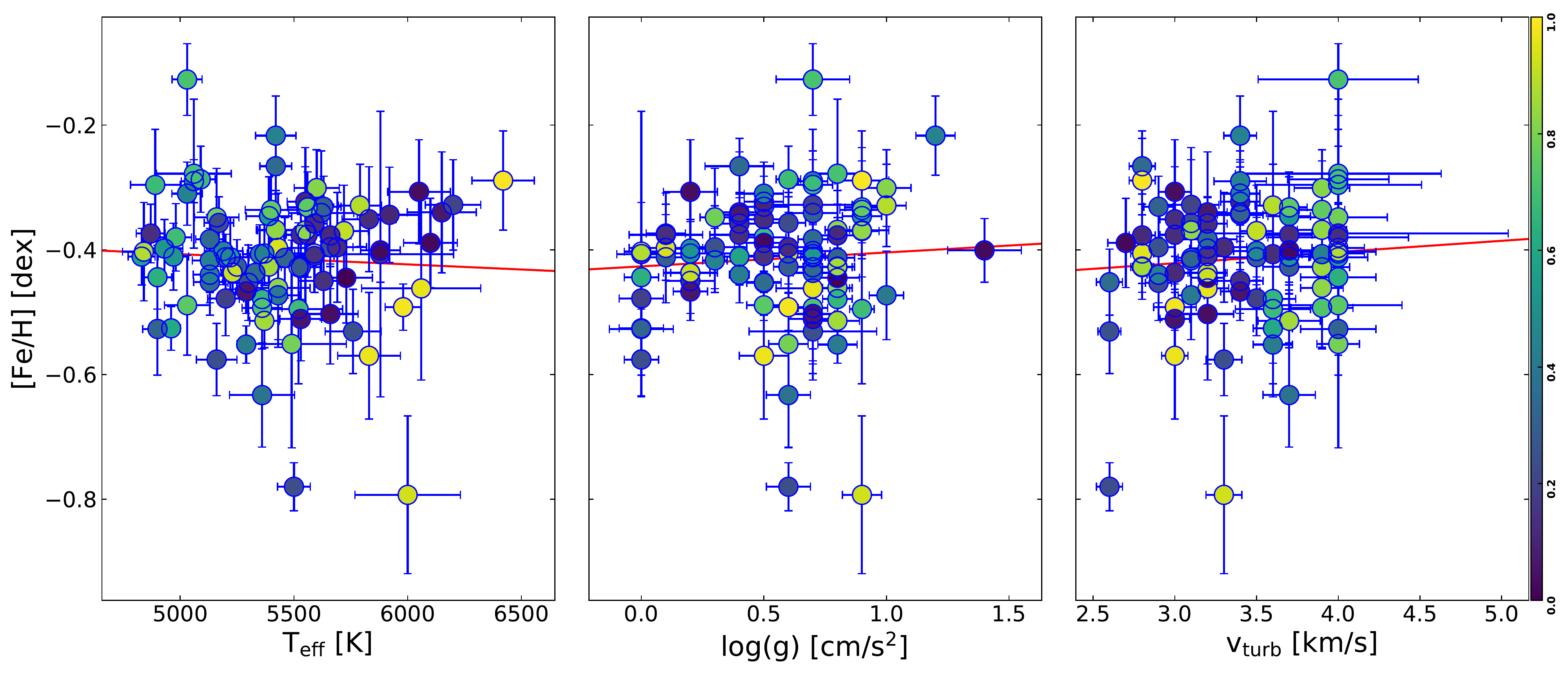}
        \caption{Measured iron abundance vs. stellar parameters simultaneously derived in the GALA analysis. As expected, hardly any residual trend is present (red line). The points are colour-coded according to the phase along the pulsational cycle within which the stars were observed, with the scale represented by the bar on the far right.}
        \label{Fig:Fe_params}
    \end{figure*}

    In order to gauge the quality of the results, in Figure~\ref{Fig:gala_diagnostics} we plot the distributions of the quantities that are minimised when computing the stellar parameters: the slope $S_\chi$ of FeI abundance versus excitation potential $\chi$ for \teff, the slope $S_\mathrm{EWR}$ of the relation between FeI abundance and reduced EW ($\mathrm{EWR)}=\log(\mathrm{EW}/\lambda)$ for \vturb, and the difference between FeI and FeII for $\logg$. In order to interpret the residuals in the convergence criteria in terms of their impact on the final abundance determination, which is our ultimate goal, in panels (a) and (b) of the same figure we also plot the corresponding peak-to-peak values of the spread in iron abundance, computed for every star as:
    
    \begin{equation}
        \Delta\mathrm{A(FeI)}_\mathrm{p2p} = S \times \mathrm{l}
        \label{Eq:p2p}
    ,\end{equation}

    where $S=S_\chi$ or $S_\mathrm{EWR}$ and $l$ is the lever range spanned in $\chi$ and $\mathrm{EWR}$, respectively, by the lines ultimately used in the fit. In other words, these peak-to-peak values are the maximum scatter in the abundance introduced by residual correlations in the minimisation process. As it can be seen, their impact is of the order of a mere few hundredths of a dex at maximum. The same applies when considering the iron abundances of the {89} stars versus the stellar parameters that, as shown in Figure~\ref{Fig:Fe_params}, do not show any appreciable residual trend (the expectation being that the chemical composition of a Cepheid does not depend on its stellar parameters along its crossings of the instability strip). The derived stellar parameters are reported in Table~\ref{Tab:parameters}
for the SH0ES sample and in Table~\ref{Tab:R08_parameters} for the archival
one. In order to make our analysis reproducible, in Appendix~\ref{App:GALA_configuration}
and \ref{App:EWs} we provide, respectively, the full GALA configuration file
we have used and the EWs of the individual lines for each star.

    \begin{table*}
        \caption[]{Stellar parameters as derived in the abundance analysis for the SH0ES sample.}
        \begin{center}
        \begin{tabular}{cccccccc}
            \hline
            CEP & \teff & d\teff & \logg & d\logg & \vturb & d\vturb & Phase\\
            \hline
            OGLE0434 & 5660 & 100 & 0.30 & 0.13 & 2.90 & 0.10 & 0.2577 \\ 
OGLE0501 & 5130 & 77 & 0.30 & 0.05 & 3.10 & 0.09 & 0.4385 \\ 
OGLE0510 & 5530 & 64 & 0.10 & 0.15 & 3.70 & 0.14 & 0.1274 \\ 
OGLE0512 & 5200 & 79 & 0.00 & 0.09 & 3.50 & 0.15 & 0.1892 \\ 
OGLE0528 & 5200 & 64 & 0.10 & 0.08 & 3.50 & 0.19 & 0.3005 \\ 
OGLE0545 & 5420 & 85 & 0.80 & 0.05 & 3.90 & 0.15 & 0.8352 \\ 
OGLE0590 & 5880 & 322 & 0.00 & 0.10 & 3.60 & 0.09 & 0.1314 \\ 
OGLE0594 & 5520 & 194 & 0.90 & 0.05 & 3.60 & 0.23 & 0.6911 \\ 
OGLE0648 & 5300 & 90 & 0.50 & 0.07 & 2.90 & 0.08 & 0.2372 \\ 
OGLE0683 & 5560 & 105 & 0.40 & 0.12 & 3.00 & 0.09 & 0.1281 \\ 
\ldots & \ldots & \ldots & \ldots & \ldots & \ldots & \ldots \\ 

            \hline
        \end{tabular}
        \end{center}
        \tablefoot{\\This table is available in its entirety in machine-readable form.\\}
        \label{Tab:parameters}
    \end{table*}
    
        \begin{table*}
        \caption[]{Stellar parameters as re-derived here in the abundance analysis for the \cite{romaniello2008} archival sample.}
        \begin{center}
        \begin{tabular}{cccccccc}
            \hline
            CEP & \teff & d\teff & \logg & d\logg & \vturb & d\vturb & Phase\\
            \hline
            OGLE107 & 5390 & 42 & 0.80 & 0.07 & 2.80 & 0.06 & 0.860 \\ 
OGLE461 & 4890 & 109 & 0.70 & 0.08 & 4.00 & 0.51 & 0.682 \\ 
OGLE655 & 5590 & 86 & 0.50 & 0.11 & 3.20 & 0.11 & 0.144 \\ 
OGLE945 & 4960 & 40 & 0.00 & 0.09 & 3.60 & 0.12 & 0.605 \\ 
OGLE1100 & 6200 & 122 & 0.70 & 0.16 & 3.10 & 0.17 & 0.201 \\ 
OGLE1128 & 5370 & 65 & 0.80 & 0.09 & 3.70 & 0.19 & 0.861 \\ 
OGLE1290 & 4870 & 64 & 0.10 & 0.08 & 4.00 & 1.04 & 0.136 \\ 
OGLE1327 & 5660 & 122 & 0.70 & 0.09 & 3.20 & 0.16 & 0.037 \\ 
OGLE2337 & 5590 & 56 & 0.40 & 0.09 & 3.20 & 0.12 & 0.130 \\ 
OGLE2832 & 5060 & 164 & 0.80 & 0.10 & 4.00 & 0.63 & 0.710 \\ 
\ldots & \ldots & \ldots & \ldots & \ldots & \ldots & \ldots \\ 

            \hline
        \end{tabular}
        \end{center}
        \tablefoot{\\This table is available in its entirety in machine-readable form.\\
        OGLE510/HV879 is common to both samples and is considered here only once, namely in Table~\ref{Tab:parameters}.}
        \label{Tab:R08_parameters}
    \end{table*}

 \subsection{Stellar versus pulsational parameters} \label{Sec:stellar_vs_pulsational}
\begin{figure}
        \centering
        \includegraphics[scale=0.4]{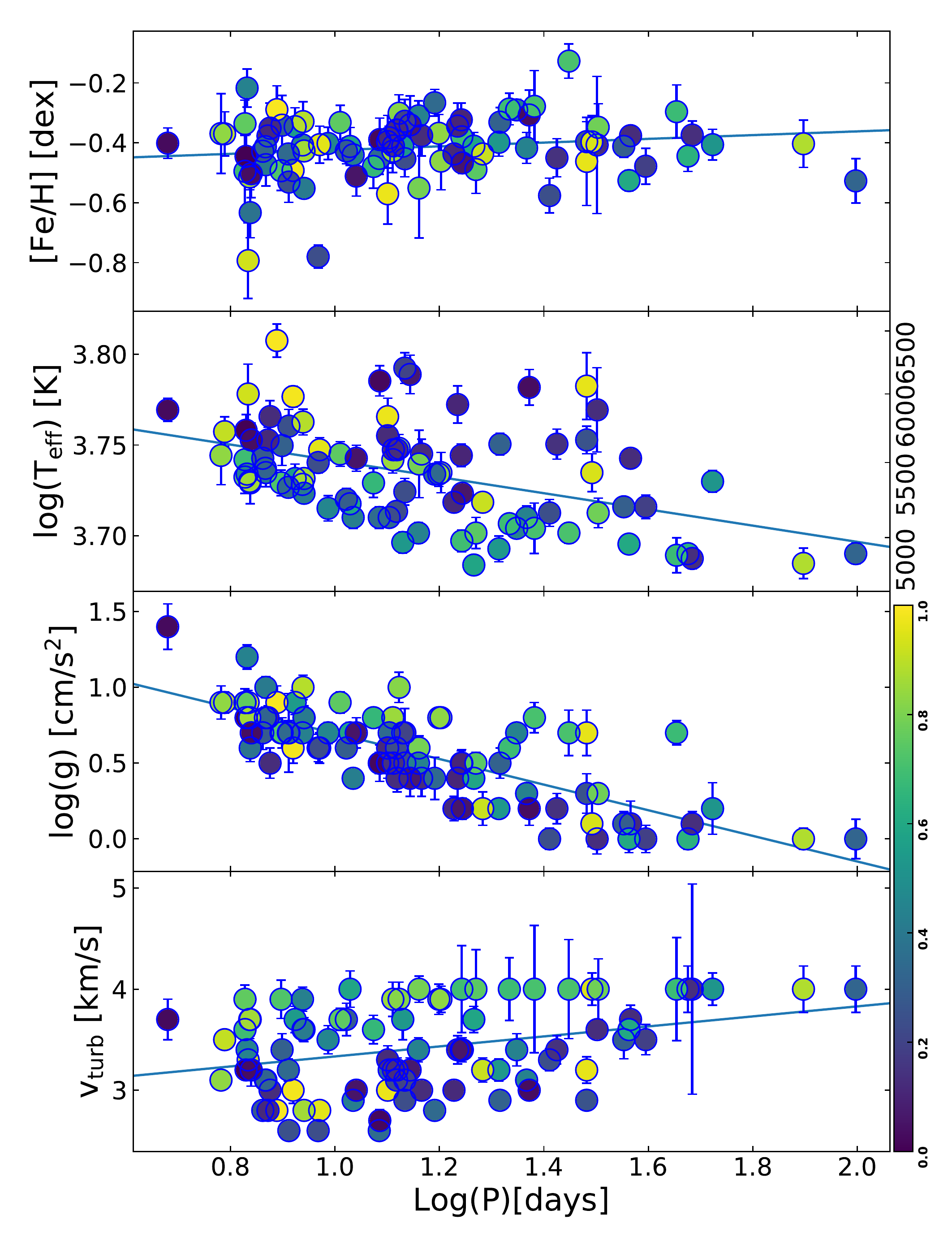}
        \caption{Stellar parameters as measured in our analysis vs the pulsational periods of the  stars. The iron abundance shows a negligible correlation with the period, further confirming the robustness of our analysis. On the other hand, for the other stellar parameters, the correlations reflect the location of the instability strip, with a scatter dominated by having observed the stars at a random phase in the pulsational cycle. The points are colour-coded according to the phase along the pulsational cycle within which the stars were observed, with the scale represented by the bar on the far right.}
        \label{Fig:logP_params}
    \end{figure}

    The fact that Cepheids pulsate according to well-defined laws allows us to perform additional consistency checks on the stellar parameters derived in the abundance analysis. To this end, in Figure~\ref{Fig:logP_params} we plot them as a function of pulsational period. As can be seen, the iron abundance shows a negligible correlation with the period, as expected from it being an intrinsic property of the star independent from the pulsation mechanism. On the other hand, correlations are found between the stellar parameters. This is also expected, in that they reflect the location of the stars in the instability strip when they were observed. The scatter is dominated by having caught the stars at a random phase along the pulsational cycle.
    
\begin{figure}
        \centering
        \includegraphics[scale=0.3]{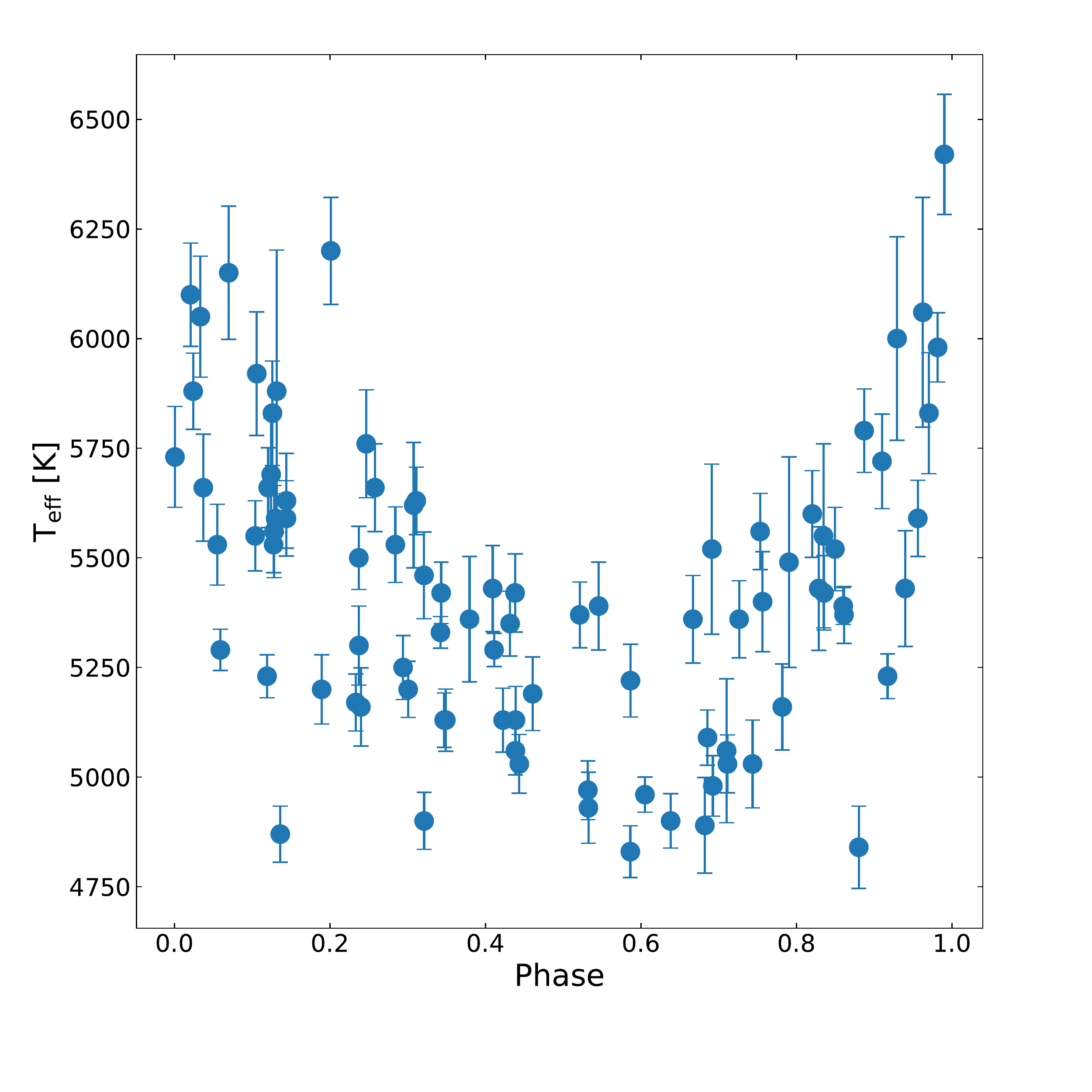}
        \caption{Derived stellar effective temperature (\teff) as a function of the phase along the pulsational cycle within which each star was observed (see Table~\ref{Tab:parameters}). The size of the mean error on \teff\ is shown by the error bar in the lower-left corner.}
        \label{Fig:phase_teff}
    \end{figure}

    Figure~\ref{Fig:phase_teff} displays the correlation of temperature with the phase along the pulsation cycle within which each star was observed. As expected, the stars are coldest towards the middle of the pulsation cycle and get hotter at either extreme, with an excursion of about 1000~K. All of these diagnostics further confirm the soundness and robustness of our analysis.

\section{Iron abundances} \label{Sec:Fe_abus}
Because of the large number of available lines in the optical spectral region, iron is often used as as a proxy for the overall metal content and as a reference against which other elemental abundances are measured. The FeI and FeII abundances derived from the procedure described above are listed in Tables~\ref{Tab:Fe_abundances} and \ref{Tab:R08_Fe_abundances} for the SH0ES and archival sample, respectively. In the following, unless otherwise noted, we
refer to FeI simply as iron when talking about abundances.

   \begin{table*}
        \caption[]{Stellar iron abundances for the SH0ES sample.}
        \begin{center}
        \begin{tabular}{ccccccccc}
            \hline
            CEP & \multicolumn{4}{c}{FeI} & \multicolumn{4}{c}{FeII} \\
                & Abundance\tablefootmark{a} & Dispersion\tablefootmark{b} & Uncertainty\tablefootmark{c} & $\mathrm{N}_\mathrm{lines}$
                & Abundance\tablefootmark{a} & Dispersion\tablefootmark{b} & Uncertainty\tablefootmark{c} & $\mathrm{N}_\mathrm{lines}$ \\
            \hline
            OGLE0434 & -0.40 & 0.12 & 0.07 & 101 & -0.43 & 0.13 & 0.04 & 15 \\ 
OGLE0501 & -0.42 & 0.09 & 0.05 & 84 & -0.43 & 0.06 & 0.03 & 12 \\ 
OGLE0510 & -0.38 & 0.11 & 0.03 & 82 & -0.39 & 0.14 & 0.03 & 9 \\ 
OGLE0512 & -0.48 & 0.13 & 0.06 & 91 & -0.41 & 0.11 & 0.01 & 14 \\ 
OGLE0528 & -0.41 & 0.12 & 0.04 & 91 & -0.43 & 0.07 & 0.02 & 15 \\ 
OGLE0545 & -0.37 & 0.09 & 0.06 & 78 & -0.40 & 0.04 & 0.05 & 10 \\ 
OGLE0590 & -0.41 & 0.10 & 0.23 & 83 & -0.38 & 0.13 & 0.08 & 15 \\ 
OGLE0594 & -0.49 & 0.14 & 0.12 & 71 & -0.50 & 0.02 & 0.10 & 6 \\ 
OGLE0648 & -0.45 & 0.09 & 0.06 & 98 & -0.47 & 0.09 & 0.04 & 17 \\ 
OGLE0683 & -0.38 & 0.11 & 0.07 & 90 & -0.40 & 0.14 & 0.05 & 16 \\ 
\ldots & \ldots & \ldots & \ldots & \ldots & \ldots & \ldots & \ldots & \ldots \\ 

            \hline
        \end{tabular}
        \end{center}
        \tablefoot{\\This table is available in its entirety in machine-readable form.\\
        Unless otherwise noted, in the following we refer to the FeI abundance simply as iron.\\
        \tablefoottext{a}{Logarithmic abundance with respect to hydrogen, in solar units: $\mathrm{[Fe/H]} \equiv A(\fe)_\mathrm{star} - A(\fe)_\sun$, where $A(\fe) = \log(N_\fe/N_\mathrm{H}) + 12$ and we adopt $A(\fe)_\sun = 7.50$ from \cite{asplund2009}.}\\
        \tablefoottext{b}{Standard deviation of the individual line measurements.}\\
        \tablefoottext{c}{Uncertainty in the abundance determination according to \cite{cayrel2004}, which includes the effects of covariance among the stellar parameters.}
        }
        \label{Tab:Fe_abundances}
    \end{table*}

   \begin{table*}
        \caption[]{Stellar iron abundances from the re-analysis of the \cite{romaniello2008} archival sample.}
        \begin{center}
        \begin{tabular}{ccccccccc}
            \hline
            CEP & \multicolumn{4}{c}{FeI} & \multicolumn{4}{c}{FeII} \\
                & Abundance\tablefootmark{a} & Dispersion\tablefootmark{b} & Uncertainty\tablefootmark{c} & $\mathrm{N}_\mathrm{lines}$
                & Abundance\tablefootmark{a} & Dispersion\tablefootmark{b} & Uncertainty\tablefootmark{c} & $\mathrm{N}_\mathrm{lines}$ \\
            \hline
            OGLE107 & -0.43 & 0.09 & 0.03 & 101 & -0.46 & 0.09 & 0.04 & 17 \\ 
OGLE461 & -0.30 & 0.15 & 0.09 & 65 & -0.30 & 0.08 & 0.10 & 9 \\ 
OGLE655 & -0.41 & 0.11 & 0.06 & 94 & -0.43 & 0.14 & 0.05 & 17 \\ 
OGLE945 & -0.53 & 0.09 & 0.04 & 83 & -0.47 & 0.09 & 0.00 & 14 \\ 
OGLE1100 & -0.33 & 0.12 & 0.07 & 79 & -0.35 & 0.19 & 0.04 & 18 \\ 
OGLE1128 & -0.51 & 0.12 & 0.04 & 82 & -0.51 & 0.10 & 0.05 & 13 \\ 
OGLE1290 & -0.37 & 0.13 & 0.05 & 59 & -0.36 & 0.07 & 0.04 & 7 \\ 
OGLE1327 & -0.50 & 0.15 & 0.08 & 81 & -0.52 & 0.10 & 0.06 & 13 \\ 
OGLE2337 & -0.36 & 0.09 & 0.04 & 87 & -0.36 & 0.11 & 0.04 & 15 \\ 
OGLE2832 & -0.28 & 0.15 & 0.12 & 63 & -0.30 & 0.10 & 0.12 & 9 \\ 
\ldots & \ldots & \ldots & \ldots & \ldots & \ldots & \ldots & \ldots & \ldots \\ 

            \hline
        \end{tabular}
        \end{center}
        \tablefoot{\\This table is available in its entirety in machine-readable form.\\
        OGLE510/HV879 is in common between the two samples and it is considered here only once, namely in Table~\ref{Tab:Fe_abundances}.\\
        Unless otherwise noted, in the following we refer to the FeI abundance simply as iron.\\
        \tablefoottext{a}{Logarithmic abundance with respect to hydrogen, in solar units: $\mathrm{[Fe/H]} \equiv A(\fe)_\mathrm{star} - A(\fe)_\sun$, where $A(\fe) = \log(N_\fe/N_\mathrm{H}) + 12$ and we adopt $A(\fe)_\sun = 7.50$ from \cite{asplund2009}.}\\
        \tablefoottext{b}{Standard deviation of the individual line measurements.}\\
        \tablefoottext{c}{Uncertainty in the abundance determination according to \cite{cayrel2004}, which includes the effects of covariance among the stellar parameters.}
        }
        \label{Tab:R08_Fe_abundances}
    \end{table*}

\begin{figure}
        \centering\includegraphics[scale=0.5]{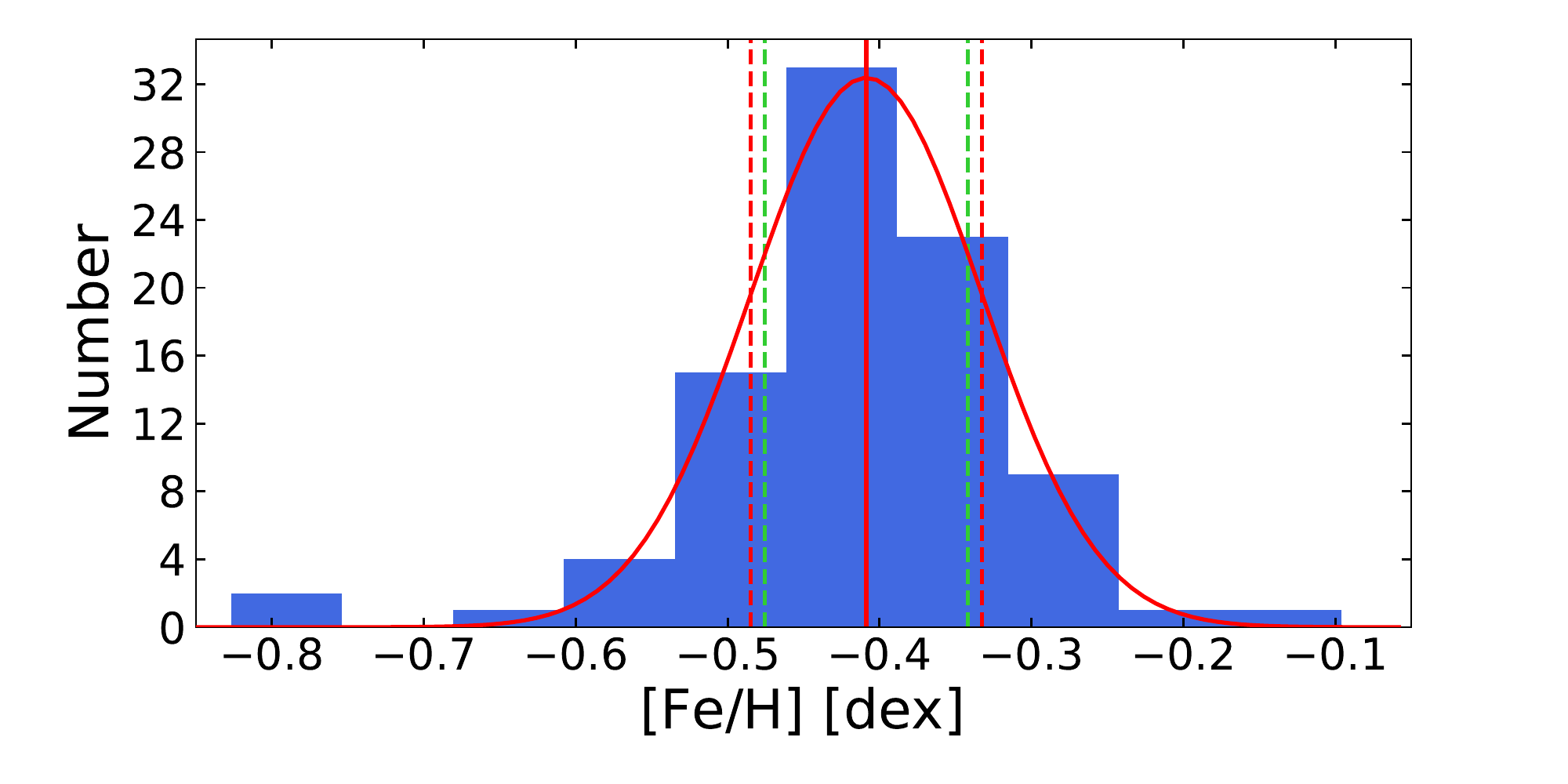}
        \caption{Histogram of the iron abundances measured for the combined sample of {89} LMC Cepheids, together with its best fitting Gaussian (solid red curve). The vertical lines mark the position of the peak of the Gaussian ($\feh=-0.409\pm0.003$, solid red), plus and minus one Gaussian sigma ($\sigma=0.076\pm0.003$, red dashed) and the mean of the error on the iron abundance resulting from the spectral analysis ($\sim0.07$~dex, green dashed line, Tables~\ref{Tab:Fe_abundances} and \ref{Tab:R08_Fe_abundances}).}
        \label{Fig:fe_hist89}
    \end{figure}

    In Figure~\ref{Fig:fe_hist89}, we show the histogram of the iron values for our 89 stars; fitting it with a Gaussian yields a mean value of $-0.409\pm0.003$~dex, with a width $\sigma=0.076\pm0.003$~dex (the values of straight mean and standard deviation are $-0.41$ and $0.09$, respectively). The latter is almost indistinguishable from the mean error as propagated through the abundance analysis ($\sim0.07$~dex, green dashed vertical lines), making the distribution consistent with a single abundance as broadened by the observational uncertainties. The uncertainty on the mean value quoted above is the random one determined by the internal consistency of our method, the intrinsic width of the distribution and the number of stars in our sample. We estimate the systematic component to be 0.1~dex (see section~\ref{Sec:fe_comparison} below). This systematic uncertainty only affects the absolute mean value of the distribution, but not its width.
    
    \subsection{Comparison with previous results} \label{Sec:fe_comparison}
    The most direct comparison with our results is with those presented by \cite{romaniello2008}. The main difference between the analyses is that here we derive the stellar effective temperature together with the other parameters (see section~\ref{Sec:params_derivation}), while \cite{romaniello2008} first fix the temperature using the Line Depth Ratio \citep[LDR; specifically in the implementation of][]{kovtyukh2000} method, and then derive the remaining parameters \logg\ and \vturb, and the iron abundance.
    
    In the left panel of Figure~\ref{Fig:fe_hist_R08}, we plot the histogram of the iron abundances from Table~9 of \cite{romaniello2008}, together with a Gaussian fit to it. The mean value of the distribution is $-0.30\pm0.02$~dex with $\sigma=0.13\pm0.02$~dex, to be compared to $-0.409$ and $0.076$~dex, respectively, for our {89} stars (see Section~\ref{Sec:Fe_abus} above).
    
    The results of the present analysis on the \cite{romaniello2008} sample are shown in the right panel of Figure~\ref{Fig:fe_hist_R08}. With a measured Gaussian mean of $-0.43\pm0.01$ and $\sigma=0.08\pm0.01$~dex, the ensemble properties of the reanalysed \cite{romaniello2008} sample compare very well with those of the {89} stars of the combined sample, as well as with those of the 68 SH0ES stars alone ($\mathrm{mean}=-0.399\pm0.003$~dex, $\sigma=0.072\pm0.003$~dex). This rules out differences between the samples and points to differences in the abundance analysis instead.
    
    The discrepancy in the mean iron abundance between the one we derive here and the one in \cite{romaniello2008} can be traced back to an offset of about 170~K in the temperatures as computed with the two methods, the present re-analysis yielding the lower values. The observed difference in iron is then consistent with the expectation that an increase in temperature of 100 K at fixed \vturb\ and \logg\ results in an increase in \feh\ of about 0.07~dex \citep{romaniello2008}.

\begin{figure*}
        \centering
        \includegraphics[scale=0.4]{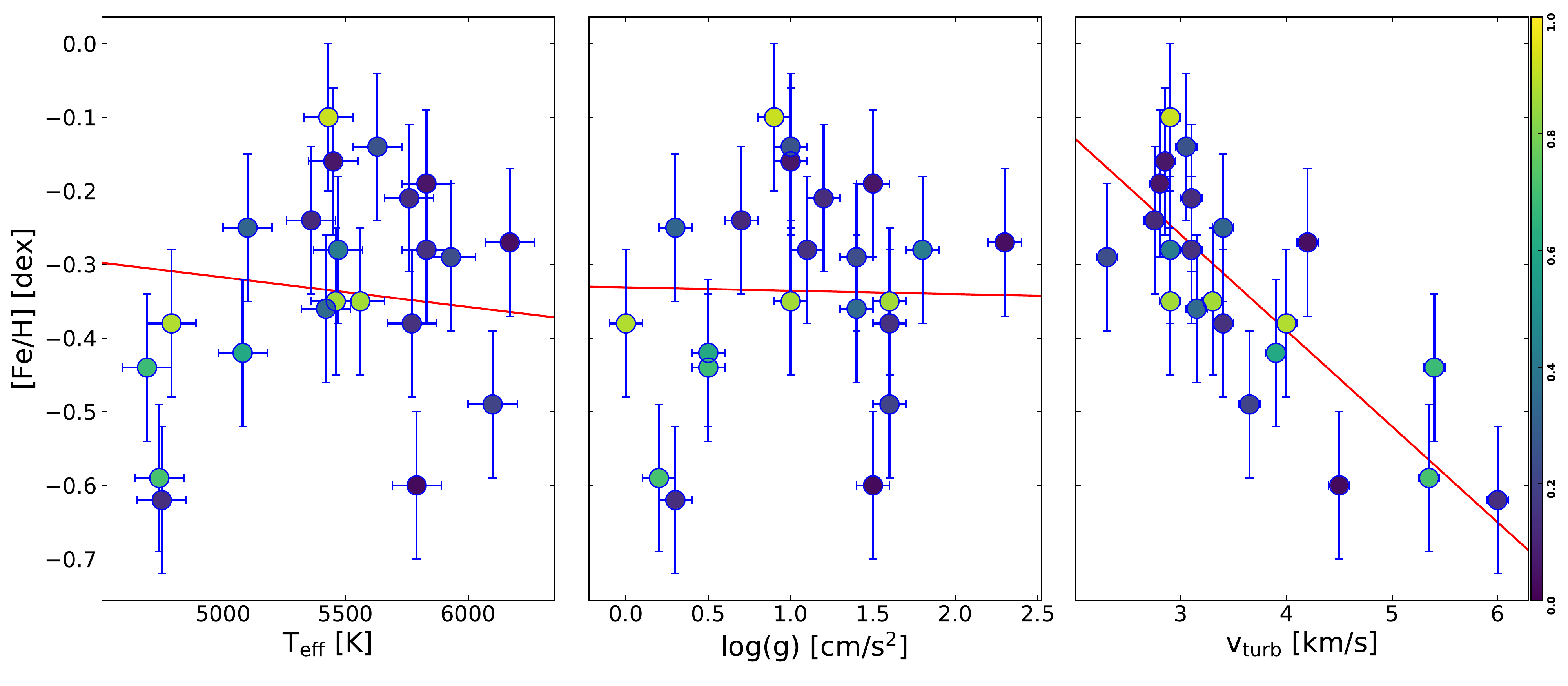}
        \caption{Measured iron abundance vs. stellar parameters as derived in \citet[][see their Table~9]{romaniello2008}, together with the results of a linear regression (red lines). A clear residual trend is present with \vturb, which is responsible for the large scatter of $\sigma=0.13$ that our analysis does not confirm. The points are colour-coded according to the phase along the pulsational cycle within which the stars were observed, with the scale represented by the bar on the far right.}
        \label{Fig:Fe_params_R08}
    \end{figure*}

\begin{figure*}
    \resizebox{\hsize}{!}
     {\includegraphics[]{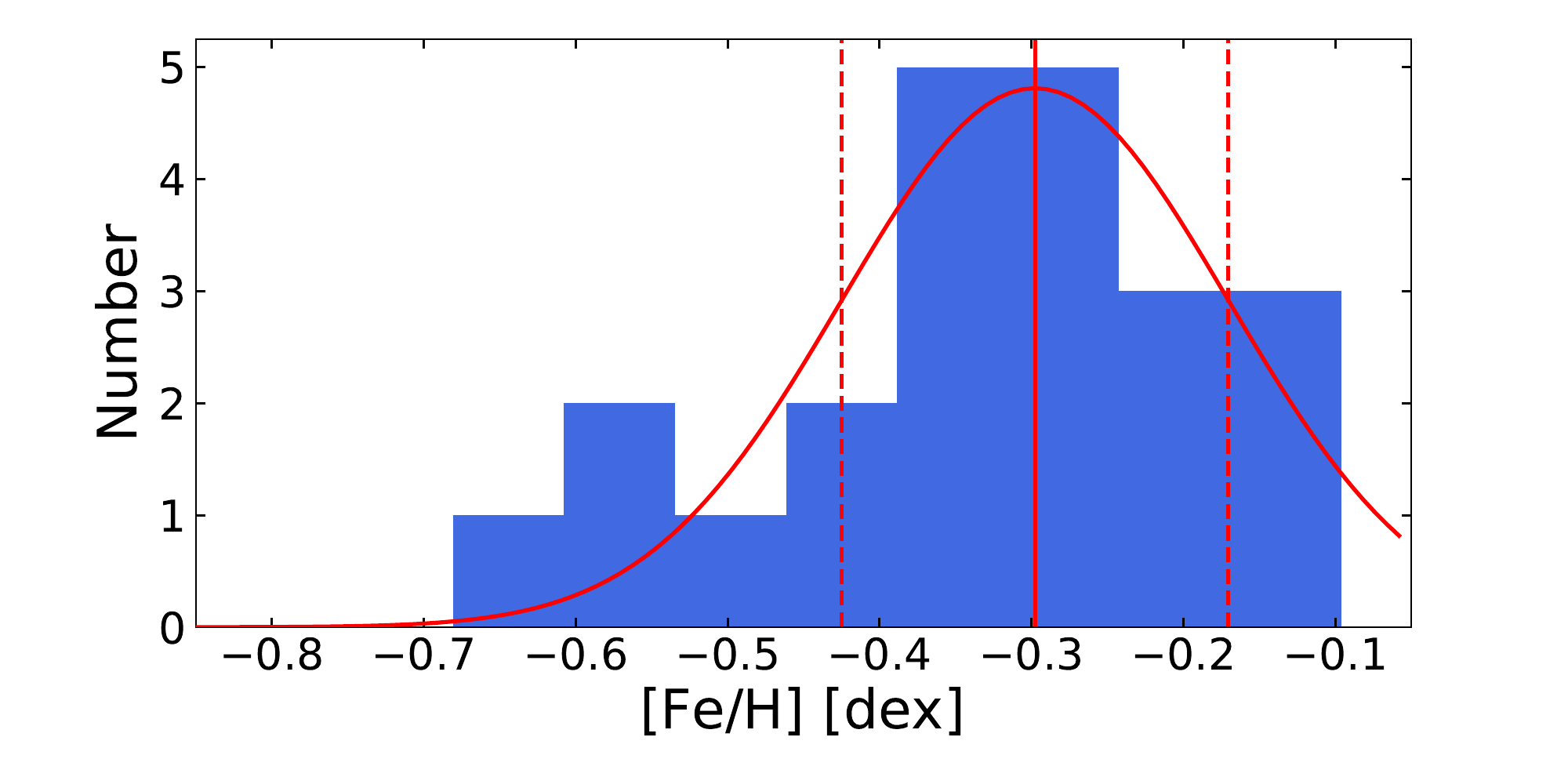}\includegraphics[]{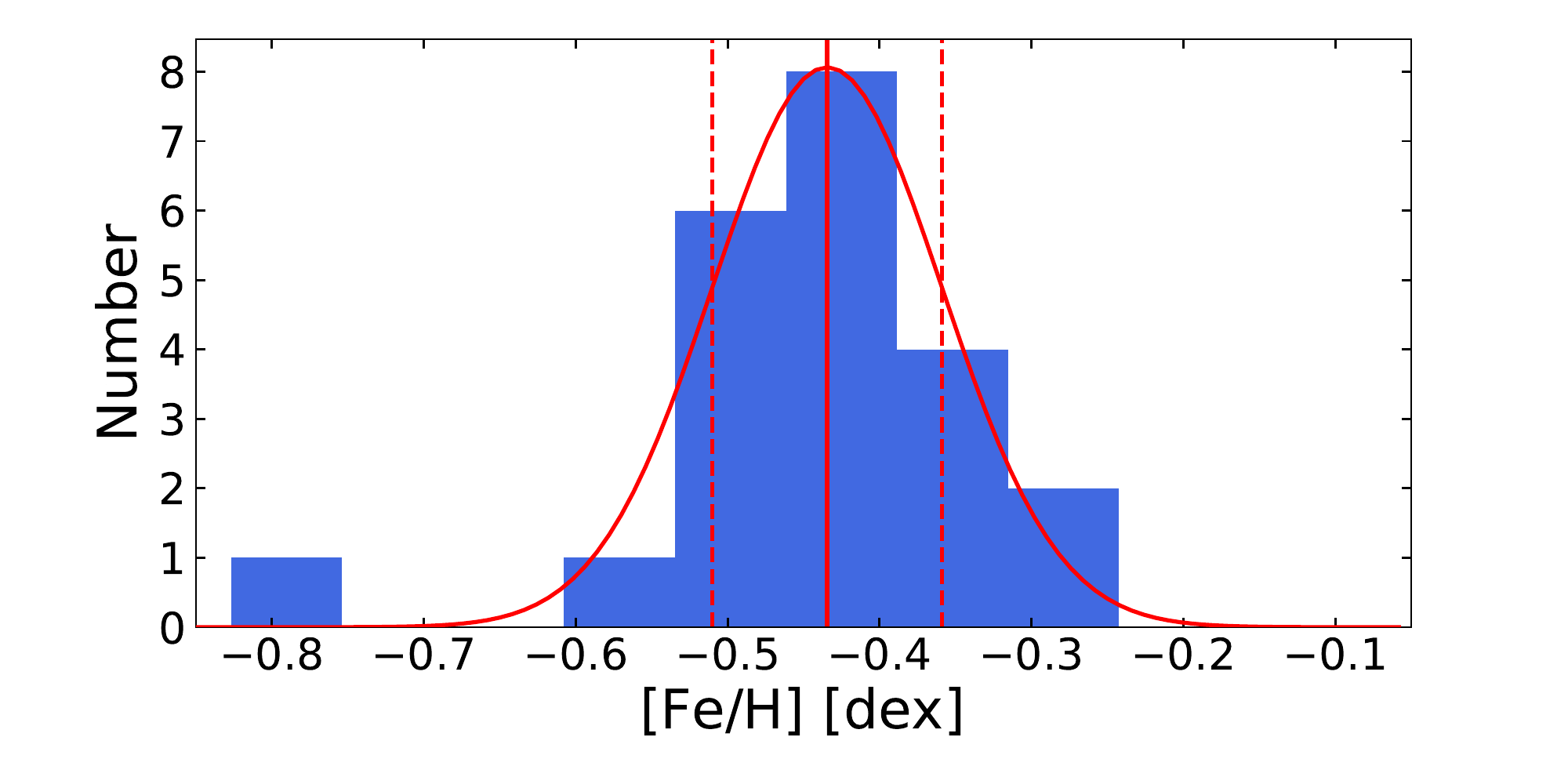}}
     \caption{Histograms of the iron abundance distributions of the sample of the 22 Cepheids in the \cite{romaniello2008} sample. {\it Left panel:} Original values from\citet[see their Table~9]{romaniello2008}, together with the best fitting Gaussian (solid red curve). The vertical lines mark the position of the peak of the Gaussian (solid red) and plus and minus one Gaussian sigma (red dashed). {\it Right panel:} Same as the left panel, but for our re-analysis of the same spectra. The distribution in the left panel was artificially broadened because of a spurious trend in \feh\ vs. \vturb\ resulting from the abundance analysis (see Figure~\ref{Fig:Fe_params_R08} and the text).}
         \label{Fig:fe_hist_R08}
    \end{figure*}

    As for the difference in dispersion, it can be explained by the fact that the analysis by \cite{romaniello2008} leaves a significant residual slope between iron abundance and stellar microturbulent velocity (see Figure~\ref{Fig:Fe_params_R08}). Once this is removed, for example with a simple linear regression, the remaining scatter is fully compatible with the quoted uncertainty in their measured iron abundances of `typically 0.08-0.1~dex'. Furthermore, in the re-analysis of the spectra, there are no significant residual trends (see Fig. 5), as is the case for the whole sample. We therefore confirm that the observed spread in iron abundances among the stars is fully compatible with a measurement uncertainty of $\sim0.08$~dex, without additional sources of broadening.
    
    Finally, we note that our results compare very well with those in \cite{urbaneja2017}, who found a mean abundance of $\feh=-0.34\pm0.11$ for 23 LMC blue supergiant stars. These stars are a different from Cepheids but the two types are coeval; they were analysed with a method that is completely independent of the one we use here, also incorporating non-LTE effects. The remarkable agreement in iron content between the two samples indicates that the effects of systematic errors in our analysis, including from possible departures from LTE, are smaller than 0.1~dex.

    \subsection{Implications for the PL relation and the distance scale} \label{Sec:fe_PLresiduals}
    The {89} LMC Cepheids we analyse here were not selected in any particular way with respect to their iron content, which was not known at all before their spectra were taken. The fact they do not show any appreciable deviation from a Gaussian with a width dictated by the rather stringent measurement uncertainty of $\sim0.08$~dex (see Figure~\ref{Fig:fe_hist89}), together with the fact that they are distributed over the full extent of the LMC (see Figure~\ref{Fig:onsky}), is indicative that it is a general property of the LMC Cepheids as a population to have the same iron abundance within that limit.
    
\begin{figure*}
        \centering
        \includegraphics[scale=0.3]{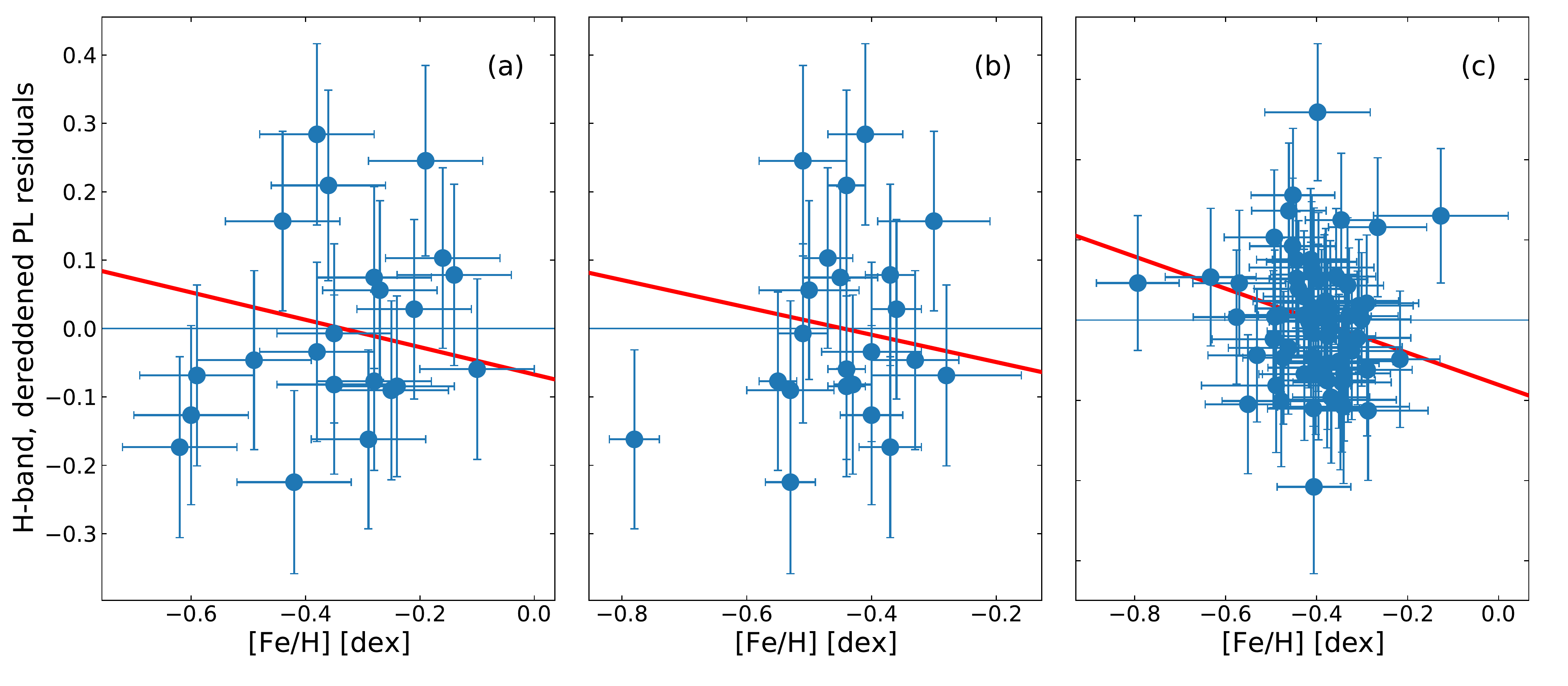}
        \caption{Residuals in the H-band dereddened magnitude PL relation as a function of iron abundance. {\it Panel~(a):} Original \cite{romaniello2008} sample of 22 LMC Cepheids, with photometry from \cite{persson2004}. {\it Panel~(b):} Same sample, but with the iron abundances as revised here. {\it Panel~(c):}  SH0ES sample of 68 LMC Cepheids, with iron abundances as measured here and photometry from \cite{riess2019}. To guide the eye, in all panels, the red line indicates a metallicity dependence of the Cepheid PL of $\gamma=-0.2$~mag/dex, which is often quoted in the recent literature \citep[e.g.][]{riess2019, breuval2021, ripepi2021}.}
        \label{Fig:h_residuals}
    \end{figure*}

    The iron content therefore does not add to the scatter of the LMC PL relation, which is advantageous when used as a tool to measure distances. On the other hand, this also means that Cepheids in the LMC alone cannot be used to determine the extent to which the relation itself depends on the chemical composition, the uncertainty on which is an important contributor to the total error budget when determining \ho\ \citep[0.9\% out of 1.8\%,][]{riess2021}. This is illustrated in the right panel of Figure~\ref{Fig:h_residuals}, where it is apparent that the span of iron abundances is too small with respect to the size of the measurement errors to determine any meaningful dependence  on the magnitude residuals in the $H$ band with respect to the fiducial PL relation of \cite{riess2019}.  For comparison, we show a line for no dependence and a $\sim-0.2$~mag/dex dependence as found by other measurements from gradients within spiral galaxies or between galaxies with difference abundances  \citep[e.g.][]{riess2019, breuval2021, ripepi2021}.  Because of the large uncertainties on both axes, any attempt to measure a dependence from this data must take care to include both in a fit.  This is inherently method-dependent for the reason that it is not well-constrained by the data and we quote illustrative results from two approaches,  $\gamma=-0.68\pm0.34$ with the {\it fitexy} algorithm described in \cite{press1992}, and $-0.11\pm1.3$ from a non-linear least-squares from a Monte-Carlo Markov Chain (lmfit).
    
      We also revisit whether $\gamma$ may be derived from the original \cite{romaniello2008} sample of 22 LMC Cepheids and their revised values here as shown in Figure~\ref{Fig:h_residuals} with important consequences for the determination of \ho.  It is apparent from Figure~\ref{Fig:h_residuals} that the measurement errors in both axes from the earlier, smaller sample are too large (and the span of [Fe/H] too small) to usefully constrain $\gamma$, with even the more optimistic {\it fitexy} algorithm giving $\sigma_\gamma > 0.4$ mag/dex for either sample.  We cannot reproduce the results of a very strong constraint of a minimal dependence of $\gamma=0.05 \pm 0.02$ for the $H$-band and  $\gamma=0.02 \pm 0.03$ for the $K$-band given by \cite{freedman2011} from  the original \cite{romaniello2008} sample which would be more than 20 times better than what we find achievable.  Not only is a small value and uncertainty in $\gamma$ not supported by the small range of \feh\ within the LMC, but these low values and uncertainties are also inconsistent with the value of $\sim-0.22 \pm 0.05$~mag/dex dependence in this region of the NIR found by \citet[][]{breuval2021} for example by comparing Cepheids between the LMC, Small Magellanic Cloud (SMC), and Milky Way to their geometric distances. Other hosts with a broader range of \feh\ can be used to measure $\gamma$ internally. This is even possible within the MW using Gaia Early Data Release 3 (EDR3) Cepheid parallaxes, where \cite{riess2021} found $\gamma=-0.20 \pm 0.12$. While comparing hosts with different mean abundance and independent geometric distances reveals a significant metallicity dependence, setting $\gamma=0$ would likewise cause the appearance of tension between the geometric anchors with different mean abundances as claimed by \cite{efstathiou2020} \footnote{Though \cite{efstathiou2020} noted that the discrepancy could be due to a metallicity dependence of the PL relation, which appears likely.}. Therefore, the inability to constrain $\gamma$ within the LMC Cepheid sample alone is an important conclusion in the context of determining the value of \ho.

\begin{figure}
        \centering
        \includegraphics[scale=0.3]{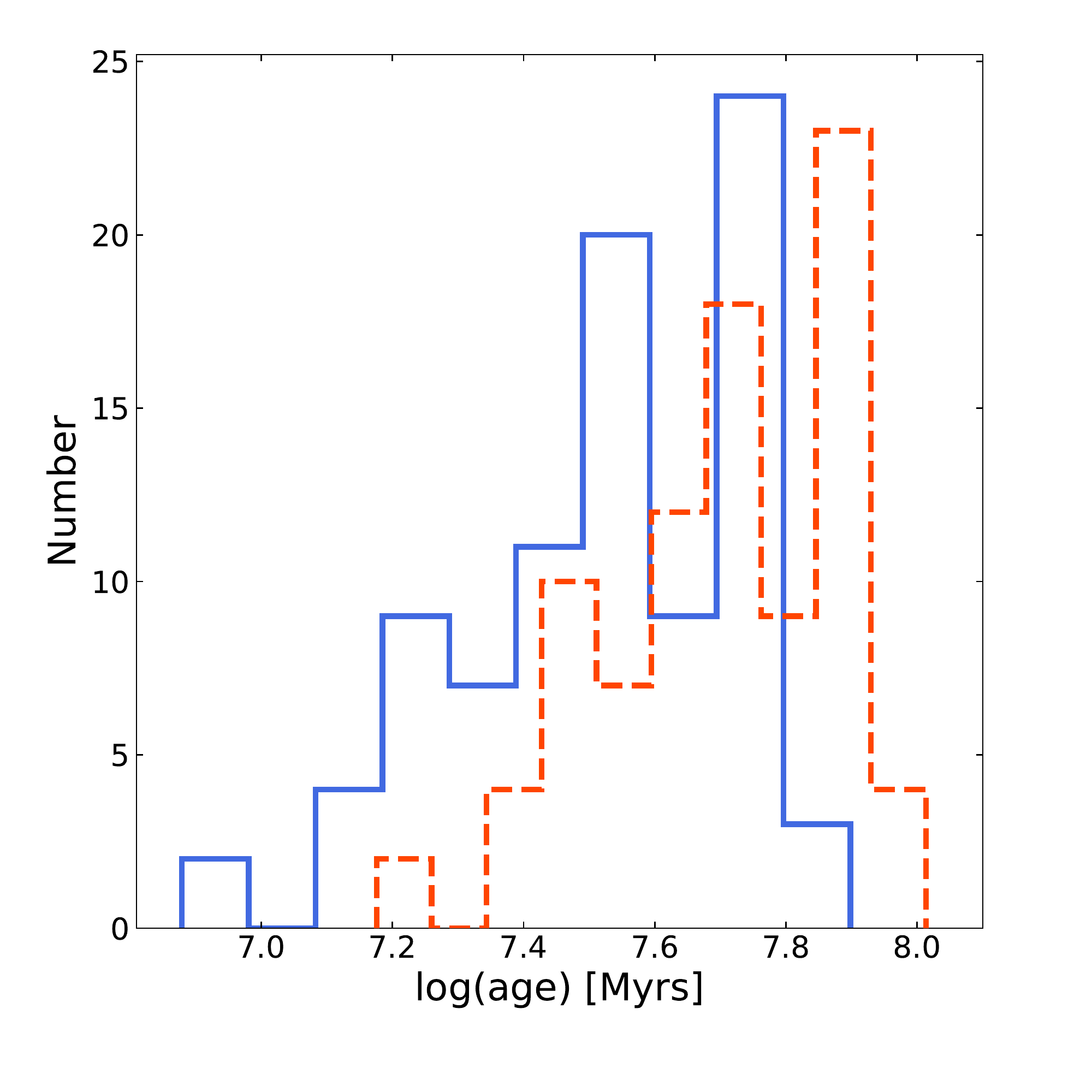}
        \caption{Age distribution of our programme stars derived by applying the period--age relations by \cite{desomma2021} for two assumptions on the amount of core convective overshooting during the core H-burning stage: no extension above the canonical value predicted by the Schwarzschild criterion (canonical scenario, solid blue line), or a moderate core overshoot (non-canonical scenario, dashed red line).}
        \label{Fig:ages}
    \end{figure}

    Such a narrow distribution in iron content across the surface of the LMC also implies that virtually no chemical enrichment took place in the look-back time covered by our sample. In order to quantify this, we converted the Cepheid pulsational periods into stellar evolutionary ages according to the relations by \cite{desomma2021}. The resulting distributions are shown in Figure~\ref{Fig:ages} for two assumptions about the amount of core convective overshooting during the core H-burning stage. Either way, the time-span is of the order of 50~Myr.
    
    \section{Oxygen abundances} \label{Sec:O_abus}
    Oxygen is the third most abundant element in the Universe after hydrogen and helium, and the most abundant one among those not created in the Big Bang. It is also fairly easy to detect and measure as strong emission lines in HII regions in galaxies along the SH0ES distance ladder to \ho\ and is therefore often used as proxy for the global metallicity \citep[e.g.][]{riess2016}. However, gas-phase measurements are only a proxy for the stellar ones and may not give consistent results \citep[e.g.][]{kewley2008, davies2017}. It is therefore important to provide direct measurements on the Cepheids themselves.
    
    However, our instrumental setup includes only very few oxygen lines from which to measure the abundance, which makes the determination rather more uncertain than that of iron. Specifically, we used the two forbidden [OI] lines at 6300.30 and 6363.77~\AA\ to measure the oxygen abundances for our stars. Both transitions cause strong airglow emissions in the night sky. However, at the spectral resolution of UVES, the heliocentric velocity at the time of observations and the radial velocity of the LMC of about 250~km/s shifts them sufficiently away from their rest-frame position so that the measurement is not affected by residuals in the sky subtraction..
    
    The results  are listed in Tables~\ref{Tab:O_abundances} and \ref{Tab:R08_O_abundances}. For 29 out of {89} stars, we could not measure a meaningful oxygen abundance, either because the EW of the lines could not be measured, or because convergence was not reached in deriving the abundance.
    
       \begin{table}
        \caption[]{Stellar oxygen abundances for the SH0ES sample.}
        \begin{center}
        \begin{tabular}{cccc}
            \hline
            CEP & \multicolumn{3}{c}{OI} \\
                & Abundance\tablefootmark{a} & Uncertainty\tablefootmark{b} &
                $\mathrm{N}_\mathrm{lines}$\tablefootmark{c} \\
            \hline
            OGLE0434 & -0.36 & 0.10 & 2 \\ 
    OGLE0501 & -0.21 & 0.10 & 2 \\ 
OGLE0510 & -0.30 & 0.06 & 1 \\ 
OGLE0512 & -0.35 & 0.06 & 2 \\ 
OGLE0528 & -0.29 & 0.07 & 2 \\ 
OGLE0545 & -0.22 & 0.10 & 1 \\ 
OGLE0590 & ---- & --- & --- \\ 
OGLE0594 & -0.12 & 0.21 & 1 \\ 
OGLE0648 & -0.32 & 0.11 & 1 \\ 
OGLE0683 & -0.28 & 0.12 & 1 \\ 
\ldots & \ldots & \ldots & \ldots \\ 

            \hline
        \end{tabular}
        \end{center}
        \tablefoot{\\This table is available in its entirety in machine-readable form.\\
        \tablefoottext{b}{Logarithmic abundance with respect to hydrogen, in solar units: $\mathrm{[O/H]} \equiv A(\ox)_\mathrm{star} - A(\ox)_\sun$, where $A(\ox) = \log(N_\ox/N_\mathrm{H}) + 12$ and we adopt $A(\ox)_\sun = 8.69$ from \cite{asplund2009}.}\\
        \tablefoottext{b}{Uncertainty in the abundance determination according to \cite{cayrel2004}, which includes the effects of covariance among the stellar parameters.} \\
        \tablefoottext{c}{No oxygen abundance is reported, either because the EW of the lines could not be measured, or because convergence was not reached in deriving the abundance (26 out of 68 stars).}
        }
        \label{Tab:O_abundances}
    \end{table}

       \begin{table}
        \caption[]{Stellar oxygen abundances from the re-analysis of the \cite{romaniello2008} archival sample.}
        \begin{center}
        \begin{tabular}{cccc}
            \hline
            CEP & \multicolumn{3}{c}{OI} \\
                & Abundance\tablefootmark{a} & Uncertainty\tablefootmark{b} &
                $\mathrm{N}_\mathrm{lines}$\tablefootmark{c} \\
            \hline
            OGLE107 & -0.33 & 0.06 & 2 \\ 
OGLE461 & 0.03 & 0.18 & 2 \\ 
OGLE655 & -0.28 & 0.10 & 1 \\ 
OGLE945 & -0.38 & 0.03 & 2 \\ 
OGLE1100 & 0.02 & 0.10 & 2 \\ 
OGLE1128 & -0.33 & 0.09 & 1 \\ 
OGLE1290 & -0.29 & 0.07 & 1 \\ 
OGLE1327 & -0.17 & 0.12 & 1 \\ 
OGLE2337 & -0.33 & 0.07 & 1 \\ 
OGLE2832 & ---- & --- & --- \\ 
\ldots & \ldots & \ldots & \ldots \\ 

            \hline
        \end{tabular}
        \end{center}
        \tablefoot{\\This table is available in its entirety in machine-readable form.\\
         OGLE510/HV879 is common to both samples and is considered here only once, namely in Table~\ref{Tab:O_abundances}.\\
        \tablefoottext{b}{Logarithmic abundance with respect to hydrogen, in solar units: $\mathrm{[O/H]} \equiv A(\ox)_\mathrm{star} - A(\ox)_\sun$, where $A(\ox) = \log(N_\ox/N_\mathrm{H}) + 12$ and we adopt $A(\ox)_\sun = 8.69$ from \cite{asplund2009}.}\\
        \tablefoottext{b}{Uncertainty in the abundance determination according to \cite{cayrel2004}, which includes the effects of covariance among the stellar parameters.} \\
        \tablefoottext{c}{No oxygen abundance is reported, either because the EW of the lines could not be measured, or because convergence was not reached in deriving the abundance (3 out of 21 stars).}
        }
        \label{Tab:R08_O_abundances}
    \end{table}

\begin{figure}
        \centering\includegraphics[scale=0.5]{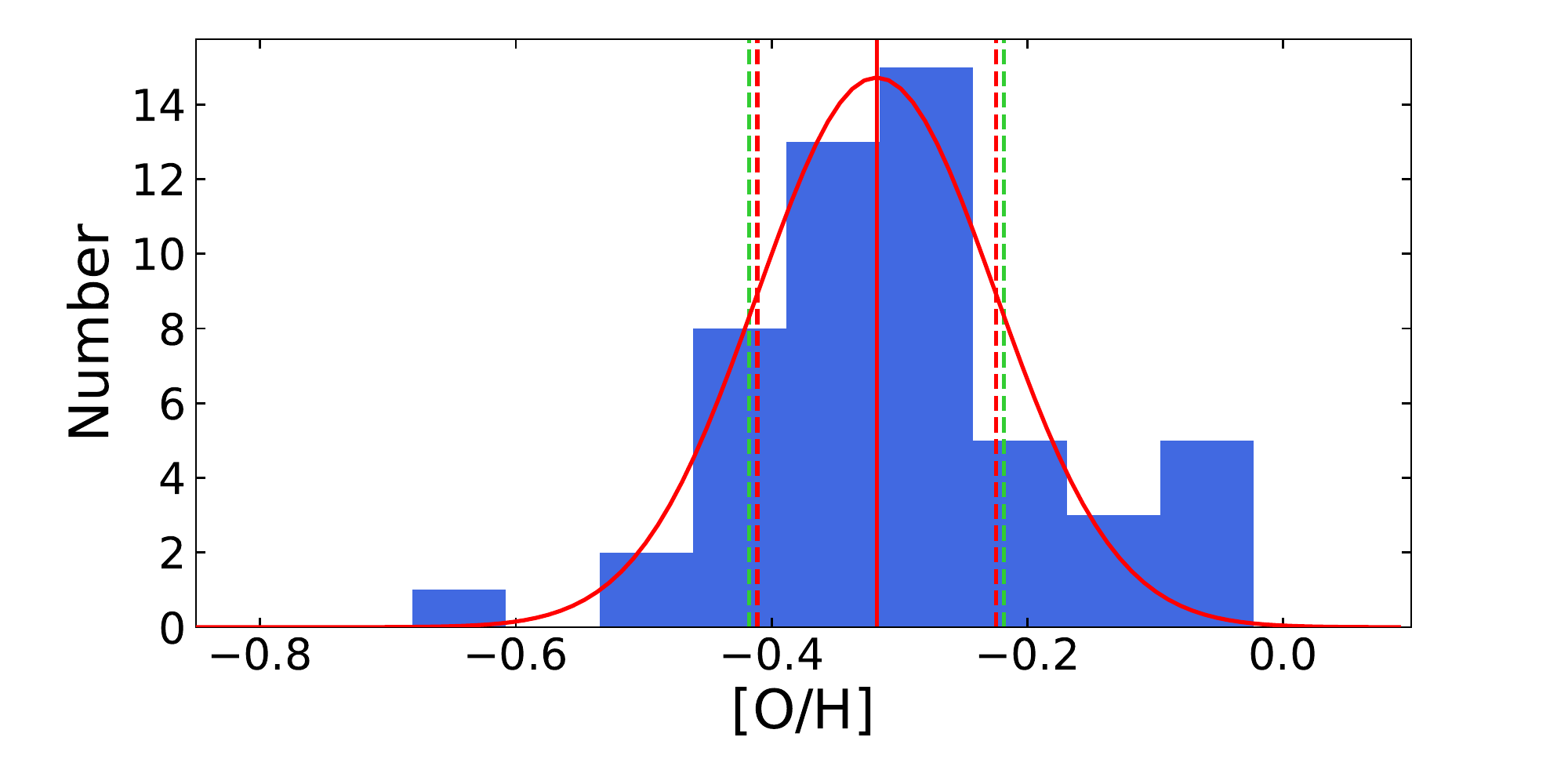}
        \caption{Histogram of the oxygen abundances measured for the 60 out of {89} LMC Cepheids (see Table~\ref{Tab:O_abundances}), together with its best fitting Gaussian (solid red curve). The vertical lines mark the position of the peak of the Gaussian (solid red), plus and minus one Gaussian sigma (red dashed) and the mean of the error on the iron abundance resulting from the spectral analysis (green dashed, Table~\ref{Tab:O_abundances}).}
        \label{Fig:o_hist}
    \end{figure}

The distribution of the measured oxygen abundances is shown in Figure~\ref{Fig:o_hist}. The mean value from a Gaussian fit is $\oxh=-0.32\pm0.01$~dex with a width of $\sigma=0.09\pm0.01$~dex, which is consistent with the mean measurement error of 0.1~dex. Therefore, the same consideration of Section~\ref{Sec:fe_PLresiduals} on the unsuitability of LMC Cepheids alone to constrain the dependence of the PL relation on iron content applies to oxygen as well.
 
    \section{Summary and conclusions} \label{Sec:conclusions}
    In this study, we analysed the spectra of {89} LMC Cepheids to measure their iron and oxygen abundances. This is of crucial relevance for the local distance scale, of which the LMC is one of the anchors.
    
    Our analysis indicates that the iron distribution of Cepheids in the LMC can be very accurately described by a single Gaussian with mean $\feh=-0.409\pm0.003$~dex and $\sigma=0.076\pm0.003$~dex for iron and 
    $\oxh=-0.32\pm0.01$~dex with a width of $\sigma=0.09\pm0.01$~dex for oxygen. We estimate a systematic uncertainty on the absolute mean values of 0.1~dex. In both cases, the observed scatter is fully compatible with the measurement error. Going beyond the present accuracy in order to characterise the chemical properties of LMC Cepheids will require a very significant effort and may not be worthwhile.
    
    Our findings do not support the earlier results by \cite{romaniello2008}, who found a significantly larger distribution in iron ($\sigma=0.13\pm0.02$~dex). The difference can be explained by a residual spurious trend of the iron abundance versus the microturbulent velocity in their analysis. The difference of 0.1~dex in the mean values of iron can be explained by the different temperature scales. Here we derive the temperatures simultaneously with the other stellar parameters and abundances, while \cite{romaniello2008} compute them a priori with the Line Depth Ratio method, also on the spectra themselves.
    
    The fact that the chemical abundance distribution is effectively unresolved given the observational uncertainties that are attainable in practice implies that the LMC alone cannot be used to constrain the possible dependency of the Cepheid luminosity on their chemical composition, which is a source of major uncertainty in measuring the Hubble constant to $\sim1$\%, as required to constrain possible deviations from the \lcdm\ Cosmological Concordance Model. The most promising way to do so seems to be to combine the LMC with the SMC and the Galaxy, so as to achieve a long enough baseline in abundance with respect to the typical uncertainties.
    
    On the other hand, the extreme chemical homogeneity of the LMC Cepheid population makes it an ideal environment in which to calibrate the PL relation at this particular metallicity. In retrospect, a crucial factor in the finding by the SH0ES team was that the period--Wesenheit relation in the HST-WFC3 H, V, and I bands has a very low intrinsic scatter of a mere 0.075~mag, or 3\% \citep{riess2019}.

\begin{acknowledgements}
    We warmly thank ESO's Observing Programme Office and User Support Department for adapting the scheduling of our observations to the evolving constraints dictated by the Covid-19 pandemic. The outstanding job of the staff of ESO's Paranal Science Operations in carrying out the observations under the gruelling circumstances is gratefully acknowledged.
    
    We thank Giulia De Somma, Marcella Marconi and Vincenzo Ripepi for fruitful discussions.
    
    We are grateful to the anonymous referee for carefully reading the manuscript and providing useful suggestions to improve it.
    
    SM and RPK acknowledge support by the Munich Excellence Cluster Origins funded by the Deutsche Forschungsgemeinschaft (DFG, German Research Foundation) under Germany's Excellence Strategy EXC-2094 390783311.
    
    RIA acknowledges support from the Swiss National Science Foundation (SNSF) through an Eccellenza Professorial Fellowship (award PCEFP2\_194638) as well as from the European Research Council (ERC) under the European Union's Horizon 2020 research and innovation programme (Grant Agreement No. 947660).
    
    This research has made use of ``Aladin sky atlas'' developed at CDS, Strasbourg Observatory, France.
\end{acknowledgements}

\begin{appendix} 
\section{GALA configuration file} \label{App:GALA_configuration}
The full set of configuration parameters we used in the abundance analysis with GALA is listed in Table~\ref{Tab:gala_autofl}. The reader is referred to the software documentation, specifically to the cookbook that comes with the software itself, for their definition and use.

  \begin{table*}
        \caption[]{Content of the GALA configuration file autofl.param.}
        \begin{center}
        \begin{tabular}{lcl}
            \hline\hline
            \multicolumn{3}{c}{Basic parameters}\\
            \hline
            \multicolumn{3}{c}{Optimisation}\\
            model    &    atlas9    & model atmospheres (atlas9,marcs...) \\
            guess    &       5      & guess working block: 0 (disabled), 1 (enabled) \\
            refine   &       1      & refine working block: 0 (disabled), 1 (enabled) \\
            uncer    &       2      & uncertainties: 0 (disabled), 1 (Cayrel), 2 (Cayrel + trad.) \\
            qiron    &   26.00      & code of the element used for the optimization \\
            epot     &       1      & option for Teff optimization (0/1) \\
            grc      &       1     & option for logg optmization (0/1/2/3) \\
            vtc      &       1      & option for vturb optimization (0/1) \\
            metal    &       1      & option for [M/H] optimization (0/1) \\
            stepgr   &     0.1      & step of logg \\
            stepteff &      30      & step of Teff (K) \\
            stepvt   &    0.1      & step of vturb (km/s) \\
            iter     &     20      & number of iterations \\
            
            \multicolumn{3}{c}{EW selection}\\
            ewstuff  &     no      & with * reads the following 3 parameters from 'list\_star'  \\
            ewmin    &   $-5.6$      & minimum log(EW/lambda) \\
            ewmax    &   $-4.5$      & maximum log(EW/lambda) \\
            error    &     15      & maximum allowed EW error (in percentage) \\
            
            \hline\hline
            \multicolumn{3}{c}{Subordinate parameters}\\
            \hline
            \multicolumn{3}{c}{Rejection}\\
            rj       &      1      & sigma-rejection from mean (0) or median (1) \\
            smax     &      3      & number of sigma used in the line rejection \\
            minl     &      3      & minimum number of lines to perform the line-rejection \\
            minnfe   &     10      & minimum number of lines to perform the optimization \\
            eplever  &    1.5      & minimum E.P. range to optimize Teff \\
            maxsig   &    1.0      & maximum abundance dispersion to perform the optimization \\
            minl     &      3      & minimum number of lines to perform the line rejection \\
            errfit   &      1      & linear fits without errors (0), errors in y (1), in x and y (2) \\
            paropt   &      0      & optimize Teff/vturb with slope (0) or Spearman coeff.(1) \\
            
            \multicolumn{3}{c}{Advanced options}\\
            kuriter  &     5      & number of blocks of 15 iterations to calculate ATLAS9 models \\
            inverse  &     -1      & values >= 0 enable the inverse analysis to derive new log(gf) \\
            int\_odf  &     1      & values >= 0 enable the interpolation in [M/H] for ATLAS9 ODFs \\
            
            \multicolumn{3}{c}{Output}\\
            plot     &      1      & produce (1) or not (0) the output plots \\
            verbose  &      3      & verbosity level (from 0 to 3) \\
            interz   &      0      & enable (1) or not (0) the interpolation to the zero value \\
            debug    &      0      & remove (0) or not (1) temporary files \\
            save     &      1      & save (1) or not (0) the new model atmospheres \\
            cog      &      1      & calculate (1) or not (0) the curve of growth for all the lines \\
            \hline
        \end{tabular}
        \end{center}
        \tablefoot{\\ The columns list, left to right, the name of the parameter, the value we have used for it and a short description.}
        \label{Tab:gala_autofl}
    \end{table*}

\section{Input equivalent widths} \label{App:EWs} 
We report here the detailed input to our abundance analysis. For each programme star, a file contains, in a format suitable for GALA, the equivalent widths of the individual spectral lines we have measured on the spectra with the DAOSPEC software, as well as the corresponding atomic physics. The columns contain:

\begin{itemize}
    \item  The wavelength (expressed in \AA).
    \item Ehe observed EW (expressed in m\AA).
    \item the uncertainty in EW (expressed in m\AA).
    \item The code of the element in the usual Kurucz notation.
    \item The logarithm of the oscillator strength.
    \item The excitation potential (expressed in eV).
    \item The logarithm of the radiative damping constant, $\gamma_{rad}$.
    \item The logarithm of the Stark damping constant, $\gamma_{stark}$.
    \item The logarithm of the Van der Waals damping constant, $\gamma_\mathrm{VdW}$.
    \item The velocity parameter $\alpha$ as defined by \cite{barklem2000}.
\end{itemize}

The reader is referred to the GALA documentation, specifically to the cookbook that comes with the software itself, for their definition and use.

\end{appendix}

\end{document}